\documentclass[aps,prb,twocolumn,superscriptaddress,showpacs]{revtex4-1}

\usepackage{graphicx}
\usepackage{amsmath}
\usepackage{float}
\usepackage{color}

\newcommand{\bsfo}{Bi$_{0.9}$Sm$_{0.1}$FeO$_3$}
\newcommand{\bfo}{BiFeO$_3$}

\begin{document}

\title{Temperature induced phase transition from cycloidal to collinear antiferromagnetism in multiferroic Bi$_{0.9}$Sm$_{0.1}$FeO$_3$ driven by $f$-$d$ induced magnetic anisotropy}

\author{R. D. Johnson}
\email{roger.johnson@physics.ox.ac.uk}
\affiliation{ISIS facility, Rutherford Appleton Laboratory-STFC, Chilton, Didcot, OX11 0QX, United Kingdom}
\affiliation{Clarendon Laboratory, Department of Physics, University of Oxford, Oxford, OX1 3PU, United Kingdom}
\author{P. A. McClarty}
\author{D. D. Khalyavin}
\author{P. Manuel}
\affiliation{ISIS facility, Rutherford Appleton Laboratory-STFC, Chilton, Didcot, OX11 0QX, United Kingdom}
\author{P. Svedlindh}
\affiliation{Department of Engineering Sciences, Uppsala University, Box 534, SE-751 21 Uppsala, Sweden}
\author{C. S. Knee}
\affiliation{Department of Chemical and Biological Engineering, Chalmers University of Technology, Gothenburg SE 412 96, Sweden}
\altaffiliation{Current address: ESAB AB, Lindholmsallén 9, SE-402 77 Gothenburg, Sweden.}

\date{\today}

\begin{abstract}
In multiferroic \bfo\ a cycloidal antiferromagnetic structure is coupled to a large electric polarization at room temperature, giving rise to magnetoelectric functionality that may be exploited in novel multiferroic-based devices. In this paper, we demonstrate that by substituting samarium for 10\% of the bismuth ions the periodicity of the room temperature cycloid is increased, and by cooling below $\sim15$ K the magnetic structure tends towards a simple G-type antiferromagnet, which is fully established at 1.5 K. We show that this transition results from $f-d$ exchange coupling, which induces a local anisotropy on the iron magnetic moments that destroys the cycloidal order - a result of general significance regarding the stability of non-collinear magnetic structures in the presence of multiple magnetic sublattices.
\end{abstract}

\maketitle

\section{Introduction}

In multiferroic materials spontaneous magnetic order is coupled to ferroelectricity such that the magnetic state may be tuned by an electric field, and \emph{vice versa} \cite{Cheong07,Johnson14}. In these technologically important materials, a \emph{polar} magnetic structure must be uniquely stabilised by the absence of inversion symmetry in an already ferroelectric phase, or itself break inversion symmetry and induce a coupled electric polarization. Such magnetic structures typically arise through a complex interplay between frustrated magnetic exchange interactions and anisotropies. Of all multiferroic materials, \bfo\ is arguably the most studied owing to its large electric polarization (over $150~\mu\mathrm{Ccm}^{-2}$ [\citenum{Yun04}]), which is coupled to antiferromagnetism at room temperature - a key requirement for technological application.

\bfo\ has a rhombohedrally-distorted perovskite crystal structure with polar space group $R3c$ \cite{Michel69, Moreau71}. The Fe$^{3+}$ magnetic sublattice is ordered at room temperature ($T_\mathrm{N}= 640~\mathrm{K}$), with a predominantly antiparallel alignment of nearest neighbor (NN) spins (G-type) resulting from dominant antiferromagnetic NN Heisenberg exchange of the form $E_\mathrm{H}=\sum_{ij}J_{ij}(\mathbf{S}_i \cdot \mathbf{S}_j)$. This energy is independent of the global direction of the magnetic moments, and in the absence of a significant Fe$^{3+}$ single ion anisotropy \cite{Jaehong12,Jaehong14} the magnetic structure of \bfo\ is stabilized by the antisymmetric Dzyaloshinskii-Moriya (DM) interaction of the form $E_\mathrm{DM} = \sum_{ij}\mathbf{D}_{ij}\cdot (\mathbf{S}_i \times \mathbf{S}_j)$ \cite{Dzyaloshinsky58,Moriya60,Katsura05,Mostovoy06}. Two distinct components of the DM vector, $\mathbf{D}$, are present in \bfo. One is allowed by global inversion-symmetry breaking of the ferroelectric order parameter. Here the component of $\mathbf{D}$ lies in the hexagonal basal plane and gains energy through a long-period cycloidal modulation of spins ($\lambda\sim$ 62 nm) within the plane containing the magnetic propagation vector $\mathbf{k}=(\delta,\delta,0)$ and $\mathbf{P}$, as observed ($\delta\sim 0.0045$) \cite{Sosnowska82,Sosnowska95,Kadomtseva04,Lebeugle08,Lee08}. The second component is allowed through antiphase FeO$_6$ octahedral rotations that are represented by an axial vector parallel to the hexagonal $c$-axis, which exactly describes the relevant component of $\mathbf{D}$ \cite{Khalyavin14}. If this interaction is finite but weaker than the first, the cycloid adopts a small orthogonal spin component whose magnitude is modulated with the same periodicity as the cycloid. If the second interaction is stronger than the first, the cycloidal state is destabilized in favour of a $|\mathbf{k}|=0$ G-type structure with a net ferromagnetic component induced by spin canting in the basal plane \cite{Ederer05}.

The crystal and magnetic structure of \bfo\ may be tuned by doping rare-earth ions onto the bismuth site. In general, lanthanide substitution reduces the average polarizability of the A-site cation, with greatest depolarization achieved with rare-earth ions of small radii \cite{Shannon76}. Increasing lanthanide content therefore destabilizes the $R3c$ polar phase, and drives the crystal structure from polar, to antipolar (reported for $Ln$ = La--Sm, and generally PbZrO$_3$-related $\sqrt{2}a_\mathrm{p} \times 4a_\mathrm{p} \times 2\sqrt{2}a_\mathrm{p}$ superstructure except for Bi$_{1-x}$La$_x$FeO$_3$, where the PbZrO$_3$-related structure is stable only in a narrow compositional range and is replaced by an incommensurate phase with $Imma(00\gamma)s00$ symmetry \cite{Rusakov11}) and finally to nonpolar GdFeO$_3$-type $\sqrt{2}a_\mathrm{p} \times 2a_\mathrm{p} \times \sqrt{2}a_\mathrm{p}$ superstructure, with each morphotropic phase transition having a profound effect on the ferroelectric and piezoelectric response of the material \cite{Troyanchuk11}. For example, holmium substitution gives rise to an enhancement in the remnant ferroelectric polarization at low switching fields \cite{Jeon11}, suppression of the spin cycloid towards a canted G-type antiferromagnetic structure was found in Bi$_{1-x}$Dy$_x$FeO$_3$ \cite{Uniyal09}, and a temperature induced 90$^\circ$ spin flop occurs upon doping neodymium \cite{Levin11}. Importantly, it has been suggested that in certain compositions ferromagnetism may coexist with ferroelectricity \cite{Khomchenko08} - critical progress towards device engineering.

At room temperature Bi$_{1-x}$Sm$_x$FeO$_3$ maintains a single $R3c$ polar phase up to a samarium content of $x\sim 0.15$. The system then transitions to a mixed $R3c$/PbZrO$_3$-related phase for compositions in the range $0.1< x < 0.18$, and finally a single non-polar GdFeO$_3$-type phase is stabilized for $x\geq 0.18$ \cite{Troyanchuk11}. Weak ferromagnetism is now well established in the non-polar phase, consistent with a symmetry allowed canting of the G-type antiferromagnetism \cite{Khomchenko10,Khomchenko_0.15_11}. Remarkably, weak ferromagnetism with a softened hysteresis has been reported in the polar phase of Bi$_{1-x}$Sm$_x$FeO$_3$ ($x<0.15$)\citep{Bielecki12,Khomchenko10}. As in undoped \bfo, ferromagnetism is incompatible with the cycloidal magnetic structure, and despite numerous studies the exact nature of this phenomenon remains unclear \cite{Nalwa08,Maurya09,Khomchenko10,Khomchenko_0.14_11,Haiyang12,Bielecki12}.

In this paper we demonstrate that the room temperature magnetic structure of \bsfo\ is indeed cycloidal, albeit with a lengthened periodicity compared to \bfo\ in accordance with a reduction in the polarization of the A-site cation sublattice. Furthermore, we show that at low temperatures the magnetic structure transitions from a spin cycloid to a simple G-type antiferromagnet, which is fully established at 1.5 K. This magnetic phase transition can be explained as a result of $f-d$ exchange coupling that induces an easy-axis magnetic anisotropy upon the Fe$^{3+}$ sublattice. This observation opens new routes for manipulating the cycloidal magnetic structure within the ferroelectric phase of \bfo.

\section{Experiment}
The polycrystalline \bsfo\ sample was synthesised via a solid state reaction route as detailed in reference \citenum{Saxin11}. Field cooled magnetization versus temperature measurement were performed between 150 and 2 K using an applied field of 5000 Oe in a Quantum Design MPMS. Neutron powder diffraction measurements were performed using the time-of-flight diffractometer WISH\cite{chapon11} at ISIS, the UK pulsed Neutron and Muon Spallation Source. 1 g of powder \bsfo\ was placed in a thin 3~mm vanadium can to minimise the effects of samarium absorption. The can was mounted inside a standard helium-4 cryostat, providing sample temperature control in the range 1.5 K to 300 K. Heat capacity data were collected from polycrystalline samples of \bfo\ and \bsfo\ as a function of temperature (cooling from 150 K to 3 K) using a Quantum Design Physical Properties Measurement System. Both samples had a mass of $\sim 5.5$ mg, and were prepared by cold pressing and sintering at 750$^\circ$C for 5 hours.

\section{Results and Discussion}

\subsection{Crystal structure}

\begin{figure}
\includegraphics[width=0.48\textwidth]{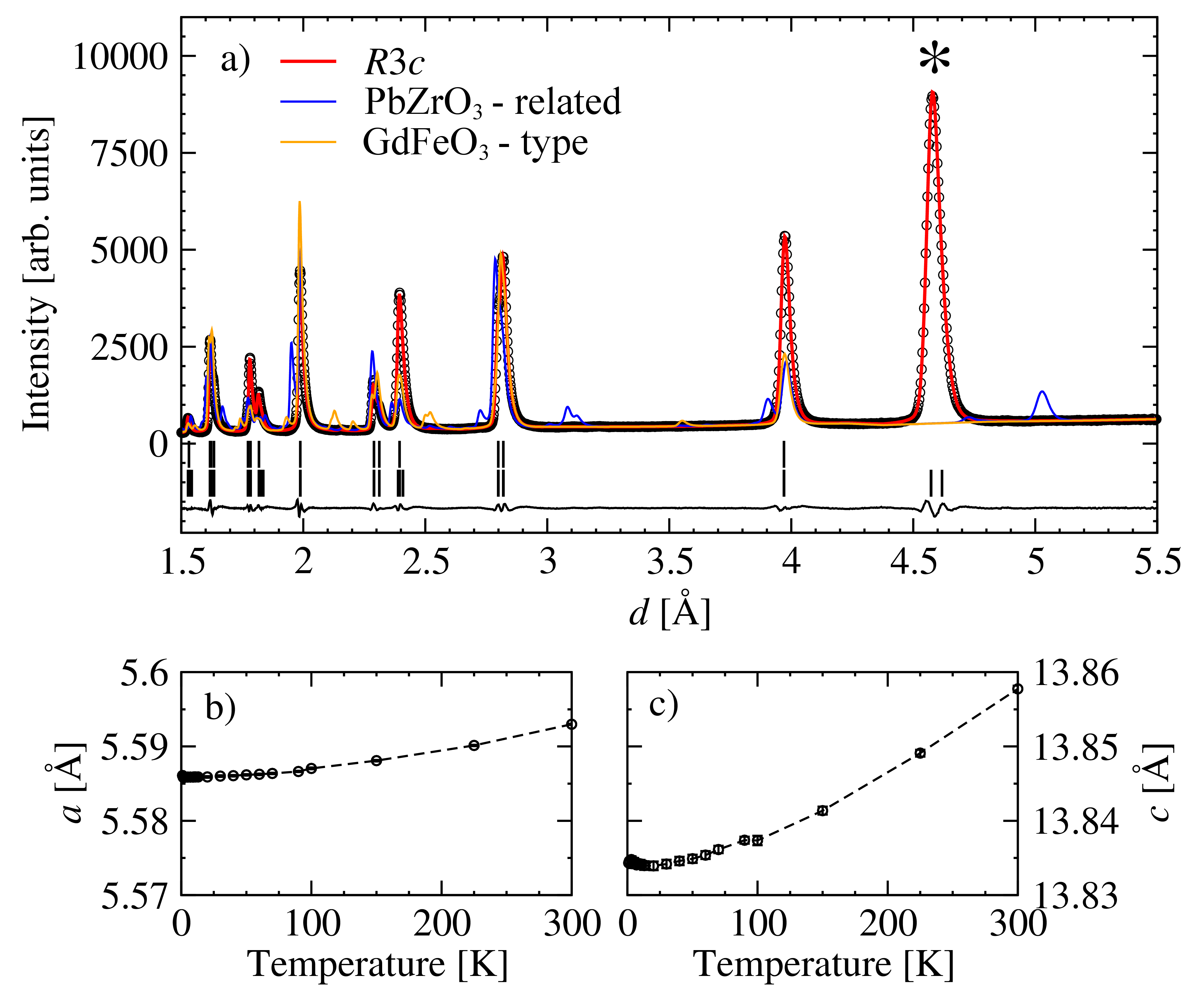}
\caption{\label{fig:npd}(Color online) a) Medium resolution neutron powder diffraction data (WISH detector bank 2) measured from \bsfo\ at 1.5 K (black circles). The polar $R3c$ crystal structure (upper tick marks and red line) is fit to the data, showing excellent agreement - the difference pattern is shown as a solid black line at the bottom of the figure. The diffraction patterns calculated for antipolar (PbZrO$_3$-related) and nonpolar (GdFeO$_3$-type) crystal structures are also shown. The magnetic diffraction peaks (lower tick marks) at $d\simeq4.57$ are labelled with an asterix. b) and c) The temperature dependence of the $a$ and $c$ lattice parameters, respectively, plotted with the same $y$-axis scale.}
\end{figure}

As \bsfo, lies close to the morphotropic phase boundary\cite{Kubota11}, and also because inhomogeneous samarium doping may lead to phase coexistence, we carried out a neutron powder diffraction study of the crystal structure of our sample. Figure \ref{fig:npd}a shows data measured at 1.5 K. These data were found to be representative of all data collected in the temperature range 1.5 to 300 K, with the exception of the magnetic diffraction peaks (discussed in detail later), and small changes in the nuclear peak positions in accordance with thermal expansion of the lattice. All nuclear diffraction peaks could be reliably fit by Rietveld refinement of the $R3c$\cite{Michel69, Moreau71} crystal structure using \textsc{fullprof}\cite{rodriguezcarvaja93} ($R_\mathrm{Bragg} = 1.57\mathrm{\%}$, $R = 6.77\mathrm{\%}$, $R_\mathrm{w} = 5.15\mathrm{\%}$). Also shown in Figure \ref{fig:npd}a, the calculated diffraction patterns for both PbZrO$_3$-related and GdFeO$_3$-type structures do not reproduce the data, and there is no qualitative evidence for phase coexistence. Hence, \bsfo\ unambiguously adopts the polar $R3c$ crystal structure at all measured temperatures (n.b. we neglect the true monoclinic symmetry of the magnetically ordered system as magneto-elastic coupling is weak). The temperature dependencies of the $a$ and $c$ lattice parameters ($R3c$ in hexagonal setting) are shown in figures \ref{fig:npd}b and \ref{fig:npd}c, respectively. The lattice was found to smoothly contract on decreasing temperature, with the largest change observed in the $c$ direction. 

\subsection{Magnetic structure}

\begin{figure}
\includegraphics[width=0.48\textwidth]{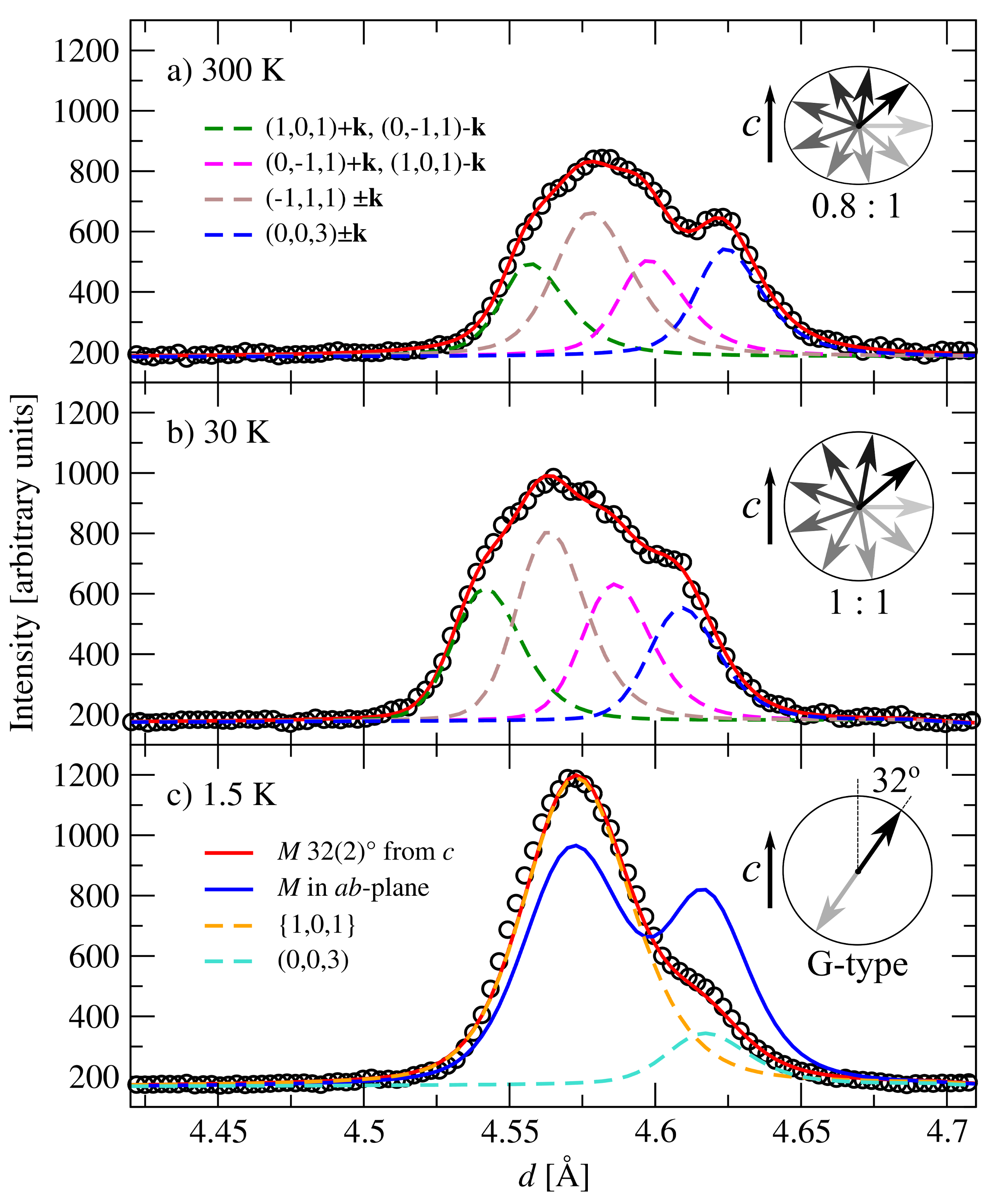}
\caption{\label{fig:peaks}(Color online) High resolution (WISH detector bank 5) magnetic neutron powder diffraction data, showing peaks labelled by an asterix in Figure \ref{fig:npd}, measured at a) 300 K, b) 30 K, and c) 1.5 K. A cartoon of the corresponding magnetic structures projected onto a single site, as found through fitting the data (solid red line), is shown on the right hand side of each pane. The solid blue line in pane c) shows the theoretical diffraction pattern for a G-type structure with in-plane magnetic moments.}
\end{figure}

The strongest magnetic diffraction peaks, labelled with an asterix in Figure \ref{fig:npd}a, were measured in high resolution at 300 K, 30 K, and 1.5 K (Figure \ref{fig:peaks}). At 300 K four reflections were observed - pairs of the six \{1,0,1\}$\pm\mathbf{k}$ satellites have common $d$-spacing forming three peaks in the powder data, and the fourth peak is a superposition of the (0,0,3)$\pm\mathbf{k}$ reflections. For the \bfo\ cycloidal magnetic structure with a circular spin rotation envelope propagating along ($\delta$, $\delta$, 0), the intensities of the (1,0,1)$\pm\mathbf{k}$ and (0,-1,1)$\pm\mathbf{k}$ peaks (green and pink dashed lines in Figure \ref{fig:peaks}) are predicted to be exactly 75\% of the (-1,1,1)$\pm\mathbf{k}$ intensity (brown dashed line), as observed. An elliptical distortion of the rotation envelope would lead to a small, almost immeasurable deviation in the relative intensities of these three peaks. The (0,0,3)$\pm\mathbf{k}$ peak intensity is dependent only on the magnitude of the $ab$-plane component of the magnetic moments, and is therefore highly sensitive to the ellipticity of the cycloid (or the degree of out-of-plane spin canting in a collinear G-type AFM phase) when measured relative to the \{1,0,1\}$\pm\mathbf{k}$ reflections.

The 300 K data in Figure \ref{fig:peaks}a was fit by Rietveld analysis using \textsc{fullprof}\cite{rodriguezcarvaja93}. The propagation vector was determined by the separation of the magnetic peaks, and refined to $\mathbf{k}=(\delta, \delta, 0)$, $\delta_\mathrm{300K} = 0.0033(1)$. The magnetic diffraction intensity was scaled to the full nuclear diffraction pattern observed at shorter $d$-spacing, giving moment magnitudes of $m_{\perp c} = 3.8(1) \mu_\mathrm{B}$ and $m_{\parallel c} = 3.1(1) \mu_\mathrm{B}$, which corresponds to an average Fe$^{3+}$ magnetic moment of 3.4(2)$\mu_\mathrm{B}$ and an ellipticity of 0.8 (defined as $m_{\parallel c} / m_{\perp c}$). The maximum moment, $m_{\perp c}$, is consistent with that found for undoped \bfo, in which $m_\mathrm{Fe} = 3.75(2)\mu_\mathrm{B}$ \cite{Sosnowska02}. The propagation vector has shortened with respect to \bfo\cite{Sosnowska82} to a value similar to that observed in Bi$_{0.9}$La$_{0.1}$FeO$_3$ ($\delta = 0.0036(5)$)\cite{Knee14}. This lengthening of the cycloidal periodicity upon doping is consistent with a depolarization of the A-site cation sublattice as the canting angle of neighboring magnetic moments is proportional to the crystal polarization. An ellipticity value $< 1$ is also consistent with a reduced polarization, as the competing DM interaction associated with octahedral rotations favors in-plane magnetic moments. 

Refinement of the cycloidal model against data measured at 30 K (Figure \ref{fig:peaks}b) showed little change in the propagation vector, $\delta_\mathrm{30K} = 0.0036(1)$, but gave moment magnitudes of $m_{\perp c} = m_{\parallel c} = 3.7(1) \mu_\mathrm{B}$ - a circular spin rotation envelope. At higher temperatures a reduction of the Fe moments when the spin lies against magnetic anisotropy may be stabilized through entropy, however, at low temperature the system will tend towards a fully saturated moment on every Fe ion.

\begin{figure}
\includegraphics[width = 0.48\textwidth]{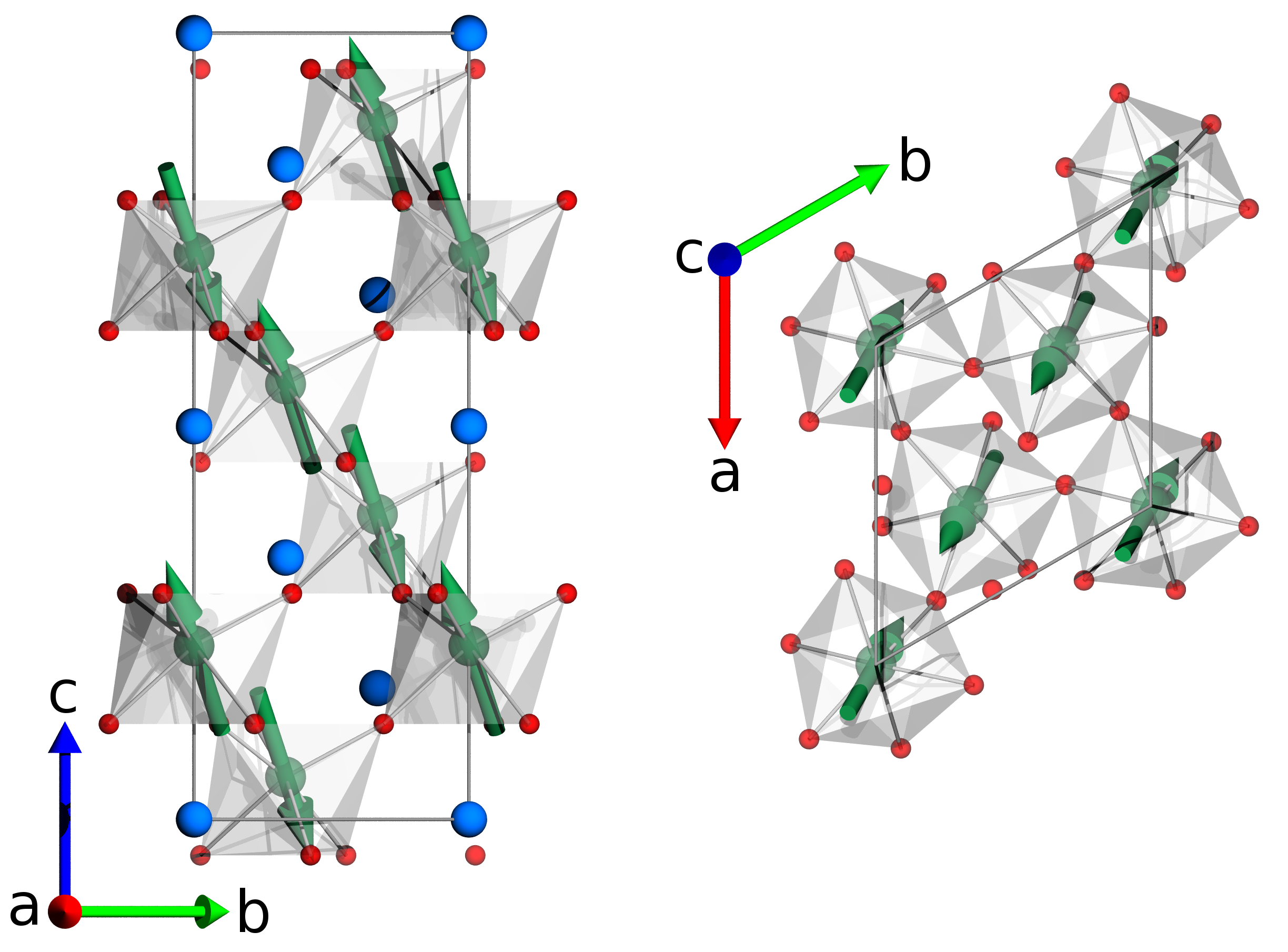}
\caption{\label{fig:magstructure}(Color online) G-type collinear antiferromagnetic structure  of \bsfo\ at 1.5 K, drawn on a single ($R3c$) unit cell. The magnetic moments (green arrows) are inclined 32(2)$^\circ$ from the $c$-axis, and drawn here to lie within the $c$-glide plane of magnetic space group $Cc$. Bismuth/samarium, iron, and oxygen ions are drawn as blue, green, and red spheres, respectively. FeO$_6$ octahedra are shaded grey.}
\end{figure}

The data measured at 1.5 K is in stark contrast to that at 30 K and 300 K. Only two peaks were observed, which index with the same families of reflections as at higher temperatures but with $|\mathbf{k}| = 0$  (\emph{i.e.} the six \{1,0,1\}$\pm\mathbf{k}$ reflections become equivalent), and corresponds to the collinear G-type antiferromagnetic structure. The G-type model was refined against the full 1.5 K data set (nuclear + magnetic) with reliability factors $R_\mathrm{nuclear} = 1.77\%$, $R_\mathrm{magnetic} = 0.31\%$, $R = 3.53\%$, and $R_\mathrm{w} = 4.60\%$. The iron moments were found to be of magnitude 3.82(3)$\mu_\mathrm{B}$ and inclined 32(2)$^\circ$ from the $c$-axis. The direction of the moment projected onto the $ab$-plane exactly determines the ground state symmetry, however, this could not be determined by powder diffraction. The magnetic space group of the cycloidal phase is $Cc1'(0,b,0)0s$, in which threefold symmetry and two $c$-glides are broken with respect to the parent symmetry, $R3c$. The incommensurate cycloid breaks translational symmetry in the hexagonal basal plane, and propagates orthogonal to the remaining glide plane. In the commensurate tilted G-type structure the threefold symmetry is also broken, as well as at least two of the three glides, but translational symmetry within the basal plane is restored. If the antiferromagnetic component of the magnetic moments lies parallel to a glide plane (\emph{i.e} in a plane orthogonal to the cycloidal rotation plane) then that glide remains and the magnetic spacegroup is $Cc$, which also allows a ferromagnetic component orthogonal to the glide plane that may couple to the axial octahedral rotations via the second DM interaction discussed in the introduction. If the antiferromagnetic component of the moments is not parallel to the glide plane the ground state symmetry is reduced to triclinic $P1$, and any moment configuration is allowed by symmetry. Whilst we cannot empirically differentiate between $Cc$ and $P1$ scenarios, we continue on the assumption that the higher symmetry structure is stablized, as drawn in Figure \ref{fig:magstructure}. 

\begin{figure}
\includegraphics[width=0.48\textwidth]{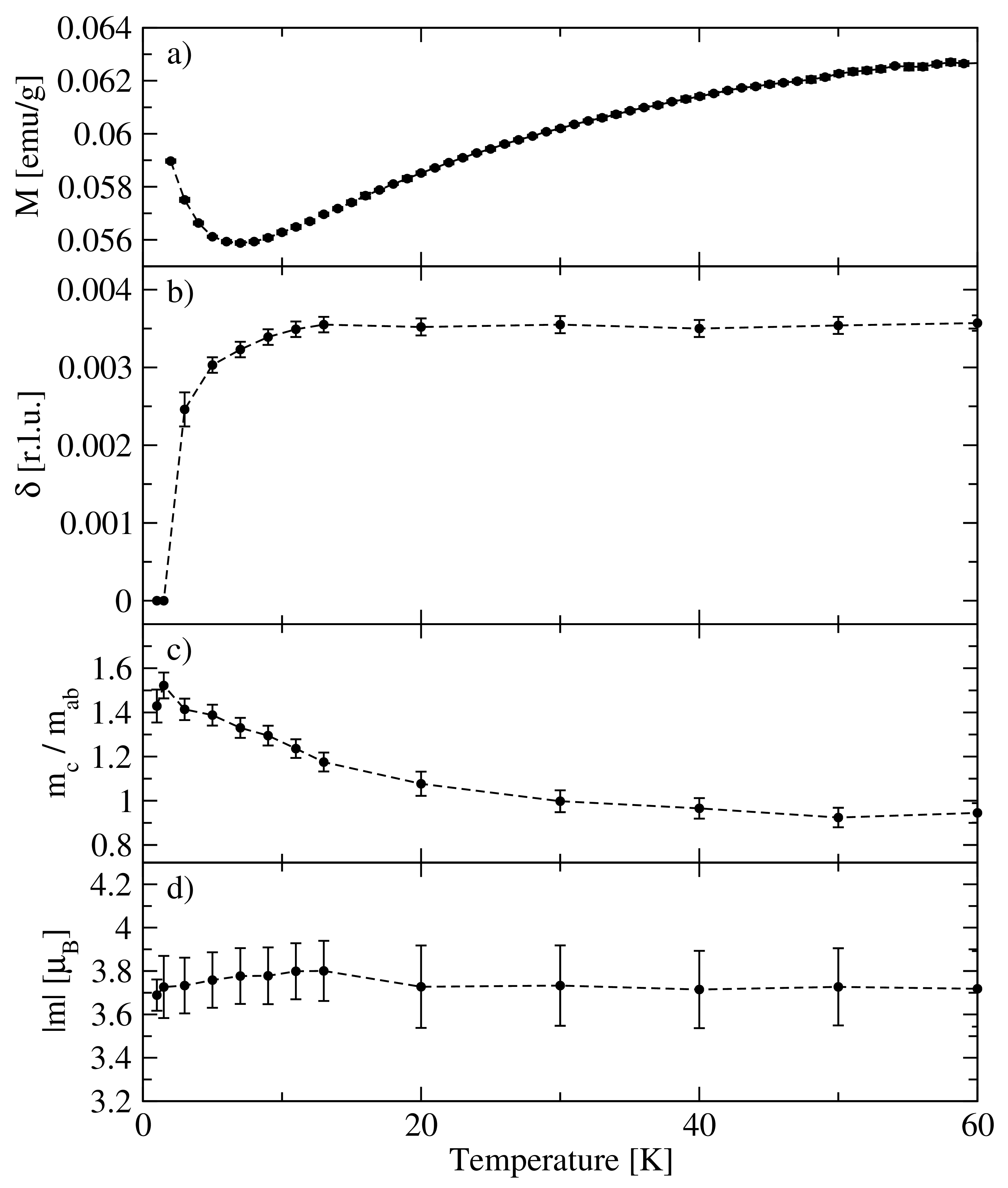}
\caption{\label{fig:magtempdep} (Color online) The temperature dependence of a) the magnetization measured on field cooling in 5000 Oe, b) the propagation vector component, $\delta$, c) the spin rotation ellipticity, and d) the average moment magnitude of the \bsfo\ magnetic structure. Note that the instrumental resolution was insufficient to determine whether or not a region of phase coexistence was present.}
\end{figure}

A phase transition from cycloidal to collinear antiferromagnetism can be observed in the temperature dependence of the magnetic propagation vector (Figure \ref{fig:magtempdep}b), which shows a trend similar to that observed for other incommensurate to commensurate magnetic phase transitions\cite{Kobayashi04}. The transition was found to be concomitant with an increase in the sample magnetisation, consistent with symmetry-allowed spin canting of the G-type magnetic structure (Figure \ref{fig:magtempdep}a). The average magnetic moment (Figure \ref{fig:magtempdep}d) was found to be invariant within error throughout the measured temperature range, whereas the ellipticity of the spin rotation envelope (Figure \ref{fig:magtempdep}c) monotonically increased on cooling below $\sim150$ K in favour of a larger magnetic moment component parallel to the hexagonal $c$-axis.

\subsection{Magnetic anisotropy and \textit{f-d} coupling}\label{MA_sec}

The origin of the temperature dependent ellipticity and the low-temperature transition to a collinear antiferromagnetic structure can be explained by the presence of a uniaxial Fe$^{3+}$ magnetic anisotropy parallel to the c-axis. If, at low temperature, the DM energy associated with octahedral rotations became larger than that of the ferroelectric polarization, a collinear magnetic structure would be favored. However, this anisotropy strictly aligns magnetic moments within the basal plane, which is incompatible with the observed change in ellipticity and the direction of the magnetic moment in the collinear phase (a structure with in-plane magnetic moments was modelled and shown as a blue line in Figure 2c). In a different Fe$^{3+}$ -based multiferroic also containing rare-earth ions, NdFe$_3$(BO$_3$)$_4$, it was demonstrated that hybridization between $4f$ and $3d$ magnetism leads to the anisotropy local to the Nd$^{3+}$ ions being inherited by the essentially isotropic Fe$^{3+}$ ions.

In orthoferrite SmFeO$_3$ (the $x = 1$ end member of the Bi$_{1-x}$Sm$_x$FeO$_3$ series), only the weak ferromagnetic component of the canted AFM iron magnetic structure couples to the rare-earth sublattice, by symmetry. However, in the rhombohedral phase of Bi$_{1-x}$Sm$_{x}$FeO$_3$, the point symmetry of the Fe$^{3+}$ site (3) allows the full AFM component of the iron magnetic moments to interact with their nearest neighbor samarium ions. With this in mind, let us consider a plausible microscopic mechanism that may lead to the observed low temperature transition. A Sm$^{3+}$ dopant ion replaces a Bi$^{3+}$ ion with a single nearest neighbor Fe$^{3+}$ separated from it by $(0,0,z)$ with $z\approx 0.22$. We consider the dominant $f$-$d$ exchange contribution between these neighboring sites. If we assume that the site symmetry of the ions is as in pure BiFeO$_3$ - namely point group $3$ - then the symmetry allowed exchange, regardless of origin, must take the form $J^x S^x + J^y S^y$, $J^z S^z$ and $J^x S^y - J^y S^x$ where $J^{\alpha}$ are the operators acting within the $J=5/2$ Sm$^{3+}$ multiplet and $S^{\alpha}$ act within the high spin ($S=5/2$) state space of the Fe$^{3+}$. The three couplings are, respectively, ${\mathcal J}_{\perp}$, ${\mathcal J}_{\parallel}$ and ${\mathcal J}_{\rm DM}$, and isotropic exchange corresponds to the case ${\mathcal J}_{\perp}={\mathcal J}_{\parallel}$ and ${\mathcal J}_{\rm DM}=0$.

At temperatures below the scale of the ground to first excited crystal field energy gap (applicable to the current study), one must consider the effective exchange, which is the ``bare" exchange described above projected onto the Sm$^{3+}$ crystal field ground state. It is therefore important to consider the possible single ion anisotropy of samarium. The site symmetry constrains the $4f$ crystal field Hamiltonian to the form $H_{\rm CF}=-\sum_{l,m} A^l_m \langle r^l_{4f} \rangle \Theta_l \hat{O}^l_m$ with $9$ potentially nonvanishing parameters: $A^2_0$, $A^4_0$, $A^{4}_{\pm 3}$, $A^{6}_{0}$, $A^{6}_{\pm 3}$,  $A^{6}_{\pm 6}$. We have written the Hamiltonian using Stevens operator equivalents which brings out a dependence on the ion-specific Stevens factors $\Theta_l$ and the radial expectation values $\langle r^l_{4f} \rangle$ of the $4f$ ions, both of which have been tabulated. For Sm$^{3+}$, $\Theta_6 = 0$, so there are four crystal field parameters for this system.

From a purely theoretical standpoint, the presence of a finite $A^2_0$ parameter means that uniaxial samarium single ion anisotropy parallel to the $c$-axis is allowed, as observed by experiment. Furthermore, if $|A^4_0|$ and $|A^{4}_{\pm 3}|$ $<$ $|A^2_0|$ [\citenum{point_charge}], the phase space spanned by the four non-zero amplitudes is dominated by an Ising-like, uniaxial state. It is therefore plausible that uniaxial anisotropy arising from the samarium crystal field forces the nearest neighbor Fe-Sm exchange to be Ising-like at low energies, giving rise to the observed low temperature transition Bi$_{0.9}$Sm$_{0.1}$FeO$_3$ (the typical scale of $4f$-$3d$ exchange is of the order $10$ K, which coincides well with the transition temperature observed experimentally). We note that the above case would result in iron magnetic moments preferentially aligning parallel to the c-axis. However, the moments were found to tilt towards the ab-plane. The tilts naturally result from the in-plane DM anisotropy acting on a collinear magnetic structure.

This result is valid in general when considering the stability of long-range-ordered cycloidal or helical magnetic structures in the presence of point like uniaxial anisotropy. For example, the spin cycloid in BiFeO$_3$ can also be destroyed by doping cobalt into the iron sites, directly introducing local magnetic anisotropy onto the transition-metal sublattice \cite{sosnowska13}. The competition between single-ion anisotropy and spin cycloids stabilized by the DM interaction in non-centrosymmetric crystals was previously discussed theoretically for isostructural PbVO$_3$ and BiCoO$_3$ \cite{Solovyev2012}. Here V$^{4+}$ has a 3d$^1$ spin configuration that forms an isotropic Kramers doublet. On the other hand, Co$^{3+}$ has a non-Kramers, anisotropic 3d$^6$ electronic configuration. As in BiFeO$_3$, ab-initio calculations predict a spin cycloid in PbVO$_3$. In contrast, single-ion anisotropy in BiCoO$_3$ partially destroys the cycloid and induces a magnetic structure of predominantly collinear AFM domains accompanied by cycloidal domain walls of width $10$ unit cells. This description is qualitatively consistent with our data measured on Bi$_{0.9}$Sm$_{0.1}$FeO$_3$ in the G-type AFM ground state, where the magnetic diffraction peaks with a Q component parallel to the magnetic propagation vector were found to be approximately 2\% broader than those measured in the cycloidal phase, which is indicative of finite size effects in the collinear AFM ground state. 

In addition to Bi$_{0.9}$Sm$_{0.1}$FeO$_3$, we have investigated the rhombohedral phases Bi$_{0.9}$Eu$_{0.1}$FeO$_3$ and Bi$_{0.9}$Tb$_{0.1}$FeO$_3$ finding that, unlike the samarium substituted case, they exhibit no low temperature transition in the magnetic structure. These results are consistent with our understanding of the mechanism driving the low temperature transition in Bi$_{0.9}$Sm$_{0.1}$FeO$_3$: Eu$^{3+}$ is nonmagnetic ($J=0$), and Tb$^{3+}$ ($J=6$) is magnetic yet it is nonKramers in a low symmetry environment so the crystal field ground state may, in principle, be a nonmagnetic singlet - both scenarios being consistent with the absence of a low temperature transition.

\subsection{Specific heat}

\begin{figure}
\includegraphics[width = 0.48\textwidth]{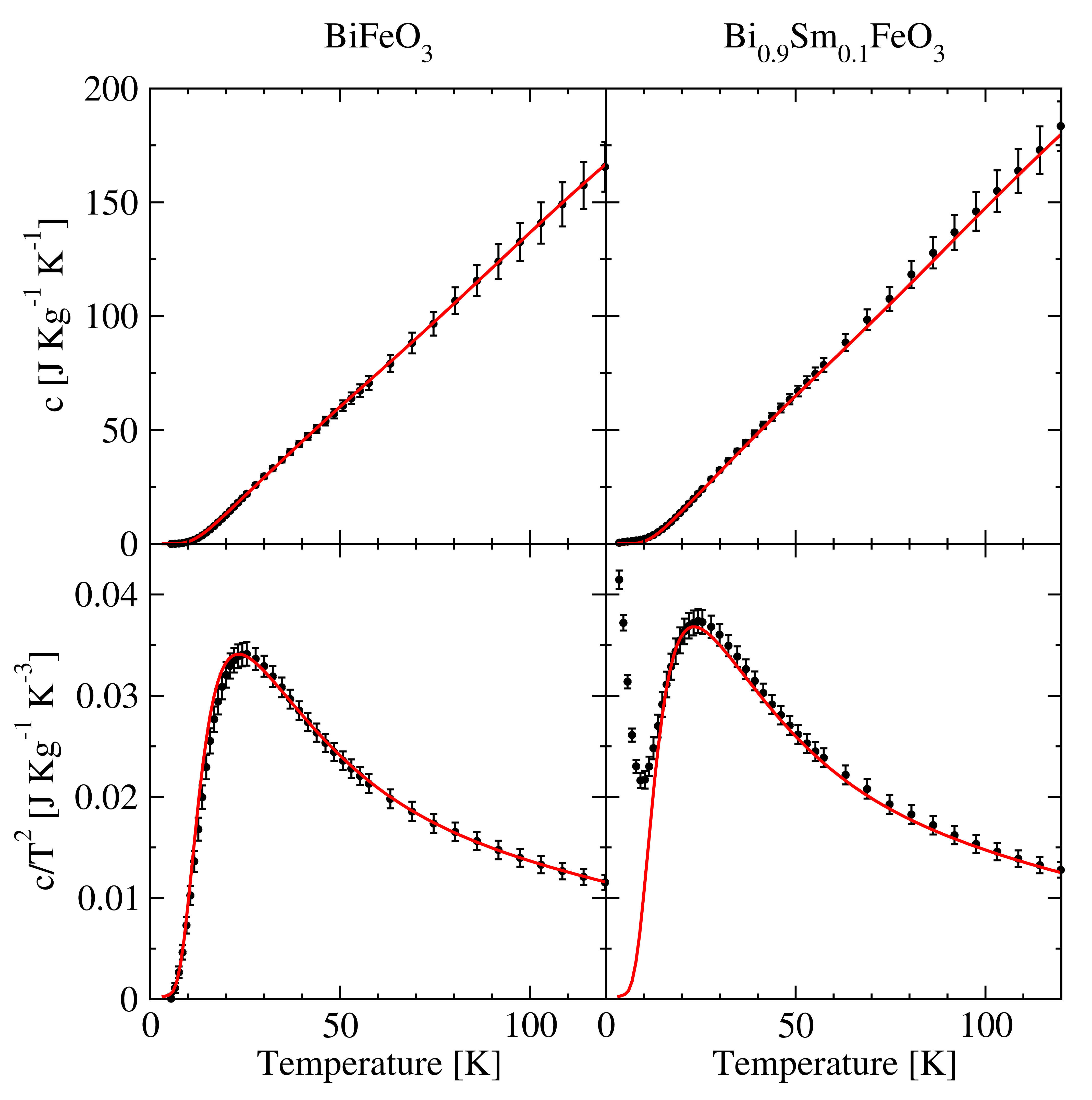}
\caption{\label{fig:HC1}(Color online) Specific heat capacity measured from both \bfo\ and \bsfo\ polycrystalline sintered pellets (black points). The bottom panes show the specific heat divided by $T^2$, in order to accentuate low temperature features. The data were fit with a Debye-Einstein model (red line).}
\end{figure}

Figure \ref{fig:HC1} shows the specific heat capacity $c(T)$ measured from both \bfo\ and \bsfo\ samples (in the lower panes $c(T)/T^2$ is plotted in order to accentuate the variation at lower temperatures). The red line shows a Debye-Einstein model simultaneously fit to both \bfo\ and \bsfo\ data sets. Here, it was assumed that phonon excitations form the dominant contribution to the specific heat of \bfo, as concluded by \emph{ab-initio} studies \cite{Wang11} (n.b. the magnon contribution to the specific heat for a collinear AFM and cycloidal magnetic structure would be expected to exhibit the same temperature dependence as the phonon contribution, except for a small magnon gap). The fitted model is in excellent quantitative agreement with the \bfo\ data. Furthermore, the number of atoms in the sample refined to $91(1)\%$ of the theoretical value, and the average frequency of the Einstein modes refined to 2.7(2) THz, which is in good agreement with the frequency of the first optical phonon mode measured by Raman scattering ($\sim2.2$ THz)\cite{Rovillain09} and found through \emph{ab-initio} calculations ($\sim2.5$ THz)\cite{Wang11}.

\begin{figure}
\includegraphics[width = 0.48\textwidth]{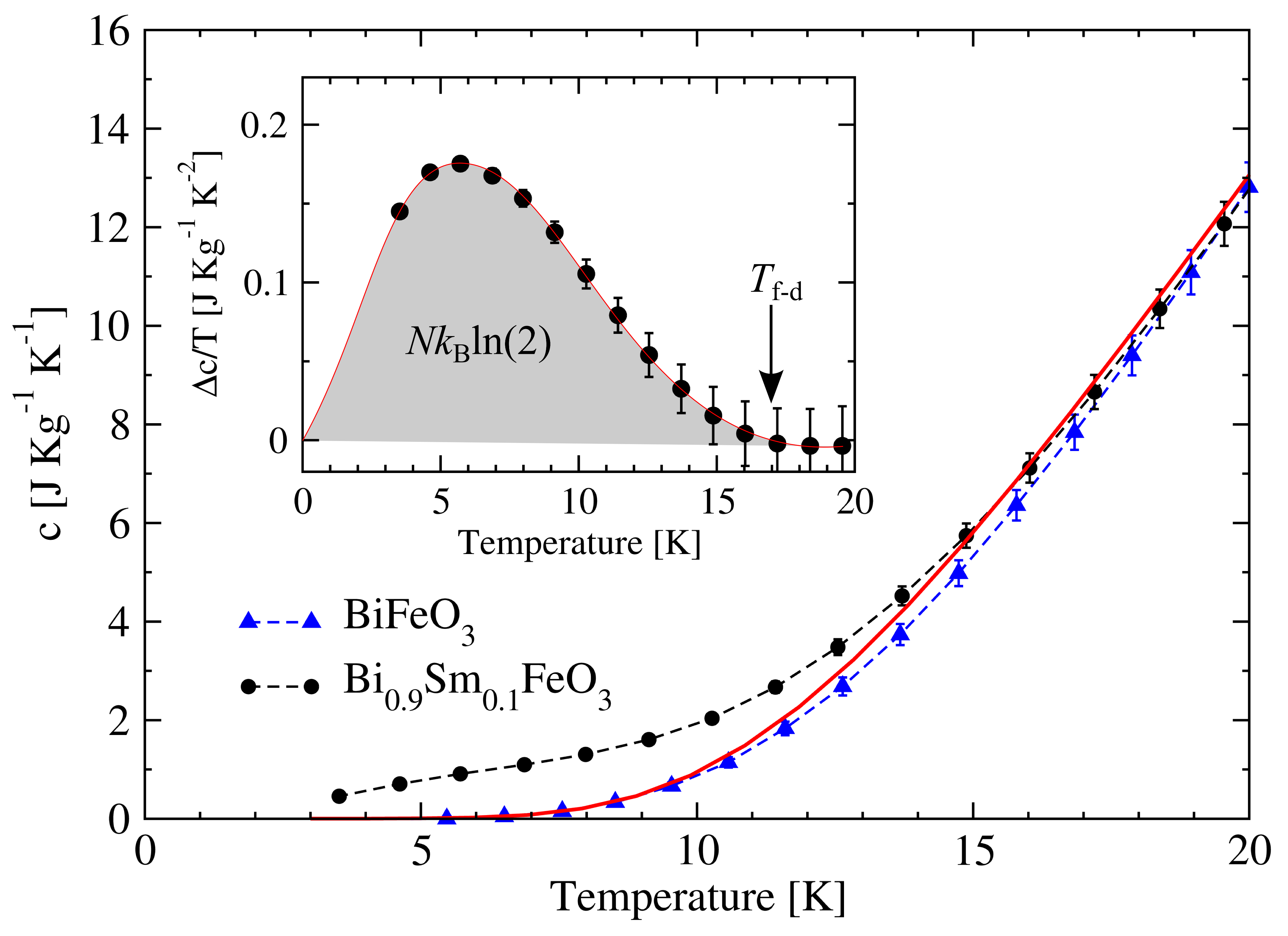}
\caption{\label{fig:HC2}(Color online) The low temperature specific heat capacity measured from both \bfo\ (blue triangles) and \bsfo\ (black circles). The inset shows the \bsfo\ data divided by temperature, having subtracted the Debye-Einstein contribution.}
\end{figure}

In \bsfo\ the specific heat was found to deviate from the Debye-Einstein model at low temperature (see Figure \ref{fig:HC1} lower-right pane and Figure \ref{fig:HC2}), indicating an additional contribution to the sample's entropy at low temperature due to the samarium substitution. In the rare-earth substituted system Nd$_{2-x}$Ce$_x$CuO$_4$, a Schottky anomaly was observed at low temperature due to a splitting of the Nd ground state doublet in the presence of \emph{f-d} exchange between Nd and Cu \cite{Brugger93}. Furthermore, at certain doping levels the low temperature side of the Schottky peak was found to exhibit a linear temperature dependence which was later found to originate in a degree of disorder within the system inherent to the doping, which effectively broadens the Nd doublet splitting \cite{Henggeler98}. These exact features were observed in our heat capacity data on \bsfo.

The inset to Figure \ref{fig:HC2} shows the anomalous contribution to the heat capacity divided by temperature, having subtracted the Debye-Einstein model. Two key parameters may be directly extracted from the data. Firstly, the onset temperature of the anomaly was found to be $\sim17$ K (labelled $T_\textit{f-d}$ in Figure \ref{fig:HC2}), which provides an indication of the \emph{f-d} exchange energy. This is in excellent agreement with the neutron diffraction results. Secondly, the integrated anomaly in $\Delta c(T)/T$ provides a measure of the change in entropy, $\Delta s$. For a two level system $\Delta s = Nk_\mathrm{B}\mathrm{ln}(2)$, where $N$ is the number of contributing atoms in the sample. Here, $N$ was found to be equal to 0.08 atoms per formula unit - in good agreement with the nominal substitution level of 0.1. Together, both quantities provide strong evidence for \emph{f-d} exchange coupling between samarium and iron magnetic moments at low temperature.

\section{Conclusions}
In summary, we have shown that \bsfo\ is isostructural with \bfo, and shares the same room temperature cycloidal magnetic structure. The real space periodicity of the cycloid was found to lengthen upon samarium substitution - consistent with the depolarization of the nominally bismuth sublattice. Furthermore, we show that at low temperatures the magnetic structure transitions from a spin cycloid to a simple G-type antiferromagnet. The G-type structure is fully established at 1.5 K, and results from an easy-axis magnetic anisotropy induced upon the Fe$^{3+}$ sublattice through $f-d$ exchange coupling. This result opens new routes for modifying the cycloidal magnetic structure of \bfo\ and, more generally, it is relevant to understanding the stability of non-collinear magnetic structures in the presence of local anisotropy that may be inherited through interactions between multiple magnetic sublattices. In future studies it would be useful to experimentally determine the crystal electric field spectrum on the samarium site using inelastic neutron scattering, in order to elucidate the fine details of the magnetic anisotropy present in this system.

\begin{acknowledgements}
RDJ acknowledges support from a Royal Society University Research Fellowship, and assistance with the heat capacity measurements from D. Prabhakaran. PAM acknowledges support from STFC through a Keeley-Rutherford fellowship. CSK acknowledges support from the Swedish research council (Vetenskapsrådet) Grant No. 621-2011-3851 and the authors thank the Science and Technology Facilities Council for provision of neutron beam time at the ISIS neutron and muon facility. DDK and PM acknowledge support from the TUMOCS project, which has received funding from the European Union's Horizon 2020 research and innovation programme under the Marie Skłodowska-Curie grant agreement No. 645660. In accordance with the UK policy framework on research data, access to the data will be made available from Ref. \citenum{ora}
\end{acknowledgements}

\bibliography{bsfo}

\begin{thebibliography}{47}%
\makeatletter
\providecommand \@ifxundefined [1]{%
 \@ifx{#1\undefined}
}%
\providecommand \@ifnum [1]{%
 \ifnum #1\expandafter \@firstoftwo
 \else \expandafter \@secondoftwo
 \fi
}%
\providecommand \@ifx [1]{%
 \ifx #1\expandafter \@firstoftwo
 \else \expandafter \@secondoftwo
 \fi
}%
\providecommand \natexlab [1]{#1}%
\providecommand \enquote  [1]{``#1''}%
\providecommand \bibnamefont  [1]{#1}%
\providecommand \bibfnamefont [1]{#1}%
\providecommand \citenamefont [1]{#1}%
\providecommand \href@noop [0]{\@secondoftwo}%
\providecommand \href [0]{\begingroup \@sanitize@url \@href}%
\providecommand \@href[1]{\@@startlink{#1}\@@href}%
\providecommand \@@href[1]{\endgroup#1\@@endlink}%
\providecommand \@sanitize@url [0]{\catcode `\\12\catcode `\$12\catcode
  `\&12\catcode `\#12\catcode `\^12\catcode `\_12\catcode `\%12\relax}%
\providecommand \@@startlink[1]{}%
\providecommand \@@endlink[0]{}%
\providecommand \url  [0]{\begingroup\@sanitize@url \@url }%
\providecommand \@url [1]{\endgroup\@href {#1}{\urlprefix }}%
\providecommand \urlprefix  [0]{URL }%
\providecommand \Eprint [0]{\href }%
\providecommand \doibase [0]{http://dx.doi.org/}%
\providecommand \selectlanguage [0]{\@gobble}%
\providecommand \bibinfo  [0]{\@secondoftwo}%
\providecommand \bibfield  [0]{\@secondoftwo}%
\providecommand \translation [1]{[#1]}%
\providecommand \BibitemOpen [0]{}%
\providecommand \bibitemStop [0]{}%
\providecommand \bibitemNoStop [0]{.\EOS\space}%
\providecommand \EOS [0]{\spacefactor3000\relax}%
\providecommand \BibitemShut  [1]{\csname bibitem#1\endcsname}%
\let\auto@bib@innerbib\@empty
\bibitem [{\citenamefont {Cheong}\ and\ \citenamefont
  {Mostovoy}(2007)}]{Cheong07}%
  \BibitemOpen
  \bibfield  {author} {\bibinfo {author} {\bibfnamefont {S.-W.}\ \bibnamefont
  {Cheong}}\ and\ \bibinfo {author} {\bibfnamefont {M.}~\bibnamefont
  {Mostovoy}},\ }\href {\doibase 10.1038/nmat1804} {\bibfield  {journal}
  {\bibinfo  {journal} {Nature Materials}\ }\textbf {\bibinfo {volume} {6}},\
  \bibinfo {pages} {13} (\bibinfo {year} {2007})}\BibitemShut {NoStop}%
\bibitem [{\citenamefont {Johnson}\ and\ \citenamefont
  {Radaelli}(2014)}]{Johnson14}%
  \BibitemOpen
  \bibfield  {author} {\bibinfo {author} {\bibfnamefont {R.~D.}\ \bibnamefont
  {Johnson}}\ and\ \bibinfo {author} {\bibfnamefont {P.~G.}\ \bibnamefont
  {Radaelli}},\ }\href {\doibase 10.1146/annurev-matsci-070813-113524}
  {\bibfield  {journal} {\bibinfo  {journal} {Annual Review of Materials
  Research}\ }\textbf {\bibinfo {volume} {44}},\ \bibinfo {pages} {269}
  (\bibinfo {year} {2014})}\BibitemShut {NoStop}%
\bibitem [{\citenamefont {Yun}\ \emph {et~al.}(2004)\citenamefont {Yun},
  \citenamefont {Ricinschi}, \citenamefont {Kanashima}, \citenamefont {Noda},\
  and\ \citenamefont {Okuyama}}]{Yun04}%
  \BibitemOpen
  \bibfield  {author} {\bibinfo {author} {\bibfnamefont {K.~Y.}\ \bibnamefont
  {Yun}}, \bibinfo {author} {\bibfnamefont {D.}~\bibnamefont {Ricinschi}},
  \bibinfo {author} {\bibfnamefont {T.}~\bibnamefont {Kanashima}}, \bibinfo
  {author} {\bibfnamefont {M.}~\bibnamefont {Noda}}, \ and\ \bibinfo {author}
  {\bibfnamefont {M.}~\bibnamefont {Okuyama}},\ }\href
  {http://stacks.iop.org/1347-4065/43/i=5A/a=L647} {\bibfield  {journal}
  {\bibinfo  {journal} {Japanese Journal of Applied Physics}\ }\textbf
  {\bibinfo {volume} {43}},\ \bibinfo {pages} {L647} (\bibinfo {year}
  {2004})}\BibitemShut {NoStop}%
\bibitem [{\citenamefont {Michel}\ \emph {et~al.}(1969)\citenamefont {Michel},
  \citenamefont {Moreau}, \citenamefont {Achenbach}, \citenamefont {Gerson},\
  and\ \citenamefont {James}}]{Michel69}%
  \BibitemOpen
  \bibfield  {author} {\bibinfo {author} {\bibfnamefont {C.}~\bibnamefont
  {Michel}}, \bibinfo {author} {\bibfnamefont {J.-M.}\ \bibnamefont {Moreau}},
  \bibinfo {author} {\bibfnamefont {G.~D.}\ \bibnamefont {Achenbach}}, \bibinfo
  {author} {\bibfnamefont {R.}~\bibnamefont {Gerson}}, \ and\ \bibinfo {author}
  {\bibfnamefont {W.}~\bibnamefont {James}},\ }\href {\doibase
  http://dx.doi.org/10.1016/0038-1098(69)90597-3} {\bibfield  {journal}
  {\bibinfo  {journal} {Solid State Communications}\ }\textbf {\bibinfo
  {volume} {7}},\ \bibinfo {pages} {701 } (\bibinfo {year} {1969})}\BibitemShut
  {NoStop}%
\bibitem [{\citenamefont {Moreau}\ \emph {et~al.}(1971)\citenamefont {Moreau},
  \citenamefont {Michel}, \citenamefont {Gerson},\ and\ \citenamefont
  {James}}]{Moreau71}%
  \BibitemOpen
  \bibfield  {author} {\bibinfo {author} {\bibfnamefont {J.}~\bibnamefont
  {Moreau}}, \bibinfo {author} {\bibfnamefont {C.}~\bibnamefont {Michel}},
  \bibinfo {author} {\bibfnamefont {R.}~\bibnamefont {Gerson}}, \ and\ \bibinfo
  {author} {\bibfnamefont {W.}~\bibnamefont {James}},\ }\href {\doibase
  http://dx.doi.org/10.1016/S0022-3697(71)80189-0} {\bibfield  {journal}
  {\bibinfo  {journal} {Journal of Physics and Chemistry of Solids}\ }\textbf
  {\bibinfo {volume} {32}},\ \bibinfo {pages} {1315 } (\bibinfo {year}
  {1971})}\BibitemShut {NoStop}%
\bibitem [{\citenamefont {Jeong}\ \emph {et~al.}(2012)\citenamefont {Jeong},
  \citenamefont {Goremychkin}, \citenamefont {Guidi}, \citenamefont {Nakajima},
  \citenamefont {Jeon}, \citenamefont {Kim}, \citenamefont {Furukawa},
  \citenamefont {Kim}, \citenamefont {Lee}, \citenamefont {Kiryukhin},
  \citenamefont {Cheong},\ and\ \citenamefont {Park}}]{Jaehong12}%
  \BibitemOpen
  \bibfield  {author} {\bibinfo {author} {\bibfnamefont {J.}~\bibnamefont
  {Jeong}}, \bibinfo {author} {\bibfnamefont {E.~A.}\ \bibnamefont
  {Goremychkin}}, \bibinfo {author} {\bibfnamefont {T.}~\bibnamefont {Guidi}},
  \bibinfo {author} {\bibfnamefont {K.}~\bibnamefont {Nakajima}}, \bibinfo
  {author} {\bibfnamefont {G.~S.}\ \bibnamefont {Jeon}}, \bibinfo {author}
  {\bibfnamefont {S.-A.}\ \bibnamefont {Kim}}, \bibinfo {author} {\bibfnamefont
  {S.}~\bibnamefont {Furukawa}}, \bibinfo {author} {\bibfnamefont {Y.~B.}\
  \bibnamefont {Kim}}, \bibinfo {author} {\bibfnamefont {S.}~\bibnamefont
  {Lee}}, \bibinfo {author} {\bibfnamefont {V.}~\bibnamefont {Kiryukhin}},
  \bibinfo {author} {\bibfnamefont {S.-W.}\ \bibnamefont {Cheong}}, \ and\
  \bibinfo {author} {\bibfnamefont {J.-G.}\ \bibnamefont {Park}},\ }\href
  {\doibase 10.1103/PhysRevLett.108.077202} {\bibfield  {journal} {\bibinfo
  {journal} {Phys. Rev. Lett.}\ }\textbf {\bibinfo {volume} {108}},\ \bibinfo
  {pages} {077202} (\bibinfo {year} {2012})}\BibitemShut {NoStop}%
\bibitem [{\citenamefont {Jeong}\ \emph {et~al.}(2014)\citenamefont {Jeong},
  \citenamefont {Le}, \citenamefont {Bourges}, \citenamefont {Petit},
  \citenamefont {Furukawa}, \citenamefont {Kim}, \citenamefont {Lee},
  \citenamefont {Cheong},\ and\ \citenamefont {Park}}]{Jaehong14}%
  \BibitemOpen
  \bibfield  {author} {\bibinfo {author} {\bibfnamefont {J.}~\bibnamefont
  {Jeong}}, \bibinfo {author} {\bibfnamefont {M.~D.}\ \bibnamefont {Le}},
  \bibinfo {author} {\bibfnamefont {P.}~\bibnamefont {Bourges}}, \bibinfo
  {author} {\bibfnamefont {S.}~\bibnamefont {Petit}}, \bibinfo {author}
  {\bibfnamefont {S.}~\bibnamefont {Furukawa}}, \bibinfo {author}
  {\bibfnamefont {S.-A.}\ \bibnamefont {Kim}}, \bibinfo {author} {\bibfnamefont
  {S.}~\bibnamefont {Lee}}, \bibinfo {author} {\bibfnamefont {S.-W.}\
  \bibnamefont {Cheong}}, \ and\ \bibinfo {author} {\bibfnamefont {J.-G.}\
  \bibnamefont {Park}},\ }\href {\doibase 10.1103/PhysRevLett.113.107202}
  {\bibfield  {journal} {\bibinfo  {journal} {Phys. Rev. Lett.}\ }\textbf
  {\bibinfo {volume} {113}},\ \bibinfo {pages} {107202} (\bibinfo {year}
  {2014})}\BibitemShut {NoStop}%
\bibitem [{\citenamefont {Dzyaloshinsky}(1958)}]{Dzyaloshinsky58}%
  \BibitemOpen
  \bibfield  {author} {\bibinfo {author} {\bibfnamefont {I.}~\bibnamefont
  {Dzyaloshinsky}},\ }\href {\doibase
  http://dx.doi.org/10.1016/0022-3697(58)90076-3} {\bibfield  {journal}
  {\bibinfo  {journal} {Journal of Physics and Chemistry of Solids}\ }\textbf
  {\bibinfo {volume} {4}},\ \bibinfo {pages} {241 } (\bibinfo {year}
  {1958})}\BibitemShut {NoStop}%
\bibitem [{\citenamefont {Moriya}(1960)}]{Moriya60}%
  \BibitemOpen
  \bibfield  {author} {\bibinfo {author} {\bibfnamefont {T.}~\bibnamefont
  {Moriya}},\ }\href {\doibase 10.1103/PhysRev.120.91} {\bibfield  {journal}
  {\bibinfo  {journal} {Phys. Rev.}\ }\textbf {\bibinfo {volume} {120}},\
  \bibinfo {pages} {91} (\bibinfo {year} {1960})}\BibitemShut {NoStop}%
\bibitem [{\citenamefont {Katsura}\ \emph {et~al.}(2005)\citenamefont
  {Katsura}, \citenamefont {Nagaosa},\ and\ \citenamefont
  {Balatsky}}]{Katsura05}%
  \BibitemOpen
  \bibfield  {author} {\bibinfo {author} {\bibfnamefont {H.}~\bibnamefont
  {Katsura}}, \bibinfo {author} {\bibfnamefont {N.}~\bibnamefont {Nagaosa}}, \
  and\ \bibinfo {author} {\bibfnamefont {A.~V.}\ \bibnamefont {Balatsky}},\
  }\href {\doibase 10.1103/PhysRevLett.95.057205} {\bibfield  {journal}
  {\bibinfo  {journal} {Phys. Rev. Lett.}\ }\textbf {\bibinfo {volume} {95}},\
  \bibinfo {pages} {057205} (\bibinfo {year} {2005})}\BibitemShut {NoStop}%
\bibitem [{\citenamefont {Mostovoy}(2006)}]{Mostovoy06}%
  \BibitemOpen
  \bibfield  {author} {\bibinfo {author} {\bibfnamefont {M.}~\bibnamefont
  {Mostovoy}},\ }\href {\doibase 10.1103/PhysRevLett.96.067601} {\bibfield
  {journal} {\bibinfo  {journal} {Phys. Rev. Lett.}\ }\textbf {\bibinfo
  {volume} {96}},\ \bibinfo {pages} {067601} (\bibinfo {year}
  {2006})}\BibitemShut {NoStop}%
\bibitem [{\citenamefont {Sosnowska}\ \emph {et~al.}(1982)\citenamefont
  {Sosnowska}, \citenamefont {Neumaier},\ and\ \citenamefont
  {Steichele}}]{Sosnowska82}%
  \BibitemOpen
  \bibfield  {author} {\bibinfo {author} {\bibfnamefont {I.}~\bibnamefont
  {Sosnowska}}, \bibinfo {author} {\bibfnamefont {T.~P.}\ \bibnamefont
  {Neumaier}}, \ and\ \bibinfo {author} {\bibfnamefont {E.}~\bibnamefont
  {Steichele}},\ }\href@noop {} {\bibfield  {journal} {\bibinfo  {journal}
  {Journal of Physics C: Solid State Physics}\ }\textbf {\bibinfo {volume}
  {15}},\ \bibinfo {pages} {4835} (\bibinfo {year} {1982})}\BibitemShut
  {NoStop}%
\bibitem [{\citenamefont {Sosnowska}\ and\ \citenamefont
  {Zvezdin}(1995)}]{Sosnowska95}%
  \BibitemOpen
  \bibfield  {author} {\bibinfo {author} {\bibfnamefont {I.}~\bibnamefont
  {Sosnowska}}\ and\ \bibinfo {author} {\bibfnamefont {A.}~\bibnamefont
  {Zvezdin}},\ }\href {\doibase http://dx.doi.org/10.1016/0304-8853(94)01120-6}
  {\bibfield  {journal} {\bibinfo  {journal} {Journal of Magnetism and Magnetic
  Materials}\ }\textbf {\bibinfo {volume} {140}},\ \bibinfo {pages} {167 }
  (\bibinfo {year} {1995})}\BibitemShut {NoStop}%
\bibitem [{\citenamefont {Kadomtseva}\ \emph {et~al.}(2004)\citenamefont
  {Kadomtseva}, \citenamefont {Zvezdin}, \citenamefont {Popov}, \citenamefont
  {Pyatakov},\ and\ \citenamefont {Vorobev}}]{Kadomtseva04}%
  \BibitemOpen
  \bibfield  {author} {\bibinfo {author} {\bibfnamefont {A.}~\bibnamefont
  {Kadomtseva}}, \bibinfo {author} {\bibfnamefont {A.}~\bibnamefont {Zvezdin}},
  \bibinfo {author} {\bibfnamefont {Y.}~\bibnamefont {Popov}}, \bibinfo
  {author} {\bibfnamefont {A.}~\bibnamefont {Pyatakov}}, \ and\ \bibinfo
  {author} {\bibfnamefont {G.}~\bibnamefont {Vorobev}},\ }\href@noop {}
  {\bibfield  {journal} {\bibinfo  {journal} {JETP Letters}\ }\textbf {\bibinfo
  {volume} {79}},\ \bibinfo {pages} {571} (\bibinfo {year} {2004})}\BibitemShut
  {NoStop}%
\bibitem [{\citenamefont {Lebeugle}\ \emph {et~al.}(2008)\citenamefont
  {Lebeugle}, \citenamefont {Colson}, \citenamefont {Forget}, \citenamefont
  {Viret}, \citenamefont {Bataille},\ and\ \citenamefont
  {Goukasov}}]{Lebeugle08}%
  \BibitemOpen
  \bibfield  {author} {\bibinfo {author} {\bibfnamefont {D.}~\bibnamefont
  {Lebeugle}}, \bibinfo {author} {\bibfnamefont {D.}~\bibnamefont {Colson}},
  \bibinfo {author} {\bibfnamefont {A.}~\bibnamefont {Forget}}, \bibinfo
  {author} {\bibfnamefont {M.}~\bibnamefont {Viret}}, \bibinfo {author}
  {\bibfnamefont {A.~M.}\ \bibnamefont {Bataille}}, \ and\ \bibinfo {author}
  {\bibfnamefont {A.}~\bibnamefont {Goukasov}},\ }\href@noop {} {\bibfield
  {journal} {\bibinfo  {journal} {Phys. Rev. Lett.}\ }\textbf {\bibinfo
  {volume} {100}},\ \bibinfo {pages} {227602} (\bibinfo {year}
  {2008})}\BibitemShut {NoStop}%
\bibitem [{\citenamefont {Lee}\ \emph {et~al.}(2008)\citenamefont {Lee},
  \citenamefont {Choi}, \citenamefont {Ratcliff}, \citenamefont {Erwin},
  \citenamefont {Cheong},\ and\ \citenamefont {Kiryukhin}}]{Lee08}%
  \BibitemOpen
  \bibfield  {author} {\bibinfo {author} {\bibfnamefont {S.}~\bibnamefont
  {Lee}}, \bibinfo {author} {\bibfnamefont {T.}~\bibnamefont {Choi}}, \bibinfo
  {author} {\bibfnamefont {W.}~\bibnamefont {Ratcliff}}, \bibinfo {author}
  {\bibfnamefont {R.}~\bibnamefont {Erwin}}, \bibinfo {author} {\bibfnamefont
  {S.-W.}\ \bibnamefont {Cheong}}, \ and\ \bibinfo {author} {\bibfnamefont
  {V.}~\bibnamefont {Kiryukhin}},\ }\href@noop {} {\bibfield  {journal}
  {\bibinfo  {journal} {Phys. Rev. B}\ }\textbf {\bibinfo {volume} {78}},\
  \bibinfo {pages} {100101} (\bibinfo {year} {2008})}\BibitemShut {NoStop}%
\bibitem [{\citenamefont {Khalyavin}\ \emph {et~al.}(2014)\citenamefont
  {Khalyavin}, \citenamefont {Salak}, \citenamefont {Olekhnovich},
  \citenamefont {Pushkarev}, \citenamefont {Radyush}, \citenamefont {Manuel},
  \citenamefont {Raevski}, \citenamefont {Zheludkevich},\ and\ \citenamefont
  {Ferreira}}]{Khalyavin14}%
  \BibitemOpen
  \bibfield  {author} {\bibinfo {author} {\bibfnamefont {D.~D.}\ \bibnamefont
  {Khalyavin}}, \bibinfo {author} {\bibfnamefont {A.~N.}\ \bibnamefont
  {Salak}}, \bibinfo {author} {\bibfnamefont {N.~M.}\ \bibnamefont
  {Olekhnovich}}, \bibinfo {author} {\bibfnamefont {A.~V.}\ \bibnamefont
  {Pushkarev}}, \bibinfo {author} {\bibfnamefont {Y.~V.}\ \bibnamefont
  {Radyush}}, \bibinfo {author} {\bibfnamefont {P.}~\bibnamefont {Manuel}},
  \bibinfo {author} {\bibfnamefont {I.~P.}\ \bibnamefont {Raevski}}, \bibinfo
  {author} {\bibfnamefont {M.~L.}\ \bibnamefont {Zheludkevich}}, \ and\
  \bibinfo {author} {\bibfnamefont {M.~G.~S.}\ \bibnamefont {Ferreira}},\
  }\href {\doibase 10.1103/PhysRevB.89.174414} {\bibfield  {journal} {\bibinfo
  {journal} {Phys. Rev. B}\ }\textbf {\bibinfo {volume} {89}},\ \bibinfo
  {pages} {174414} (\bibinfo {year} {2014})}\BibitemShut {NoStop}%
\bibitem [{\citenamefont {Ederer}\ and\ \citenamefont
  {Spaldin}(2005)}]{Ederer05}%
  \BibitemOpen
  \bibfield  {author} {\bibinfo {author} {\bibfnamefont {C.}~\bibnamefont
  {Ederer}}\ and\ \bibinfo {author} {\bibfnamefont {N.~A.}\ \bibnamefont
  {Spaldin}},\ }\href {\doibase 10.1103/PhysRevB.71.060401} {\bibfield
  {journal} {\bibinfo  {journal} {Phys. Rev. B}\ }\textbf {\bibinfo {volume}
  {71}},\ \bibinfo {pages} {060401} (\bibinfo {year} {2005})}\BibitemShut
  {NoStop}%
\bibitem [{\citenamefont {Shannon}(1976)}]{Shannon76}%
  \BibitemOpen
  \bibfield  {author} {\bibinfo {author} {\bibfnamefont {R.~D.}\ \bibnamefont
  {Shannon}},\ }\href {\doibase 10.1107/S0567739476001551} {\bibfield
  {journal} {\bibinfo  {journal} {Acta Crystallographica Section A}\ }\textbf
  {\bibinfo {volume} {32}},\ \bibinfo {pages} {751} (\bibinfo {year}
  {1976})}\BibitemShut {NoStop}%
\bibitem [{\citenamefont {Rusakov}\ \emph {et~al.}(2011)\citenamefont
  {Rusakov}, \citenamefont {Abakumov}, \citenamefont {Yamaura}, \citenamefont
  {Belik}, \citenamefont {Van~Tendeloo},\ and\ \citenamefont
  {Takayama-Muromachi}}]{Rusakov11}%
  \BibitemOpen
  \bibfield  {author} {\bibinfo {author} {\bibfnamefont {D.~A.}\ \bibnamefont
  {Rusakov}}, \bibinfo {author} {\bibfnamefont {A.~M.}\ \bibnamefont
  {Abakumov}}, \bibinfo {author} {\bibfnamefont {K.}~\bibnamefont {Yamaura}},
  \bibinfo {author} {\bibfnamefont {A.~A.}\ \bibnamefont {Belik}}, \bibinfo
  {author} {\bibfnamefont {G.}~\bibnamefont {Van~Tendeloo}}, \ and\ \bibinfo
  {author} {\bibfnamefont {E.}~\bibnamefont {Takayama-Muromachi}},\ }\href
  {\doibase 10.1021/cm1030975} {\bibfield  {journal} {\bibinfo  {journal}
  {Chemistry of Materials}\ }\textbf {\bibinfo {volume} {23}},\ \bibinfo
  {pages} {285} (\bibinfo {year} {2011})}\BibitemShut {NoStop}%
\bibitem [{\citenamefont {Troyanchuk}\ \emph {et~al.}(2011)\citenamefont
  {Troyanchuk}, \citenamefont {Karpinsky}, \citenamefont {Bushinsky},
  \citenamefont {Mantytskaya}, \citenamefont {Tereshko},\ and\ \citenamefont
  {Shut}}]{Troyanchuk11}%
  \BibitemOpen
  \bibfield  {author} {\bibinfo {author} {\bibfnamefont {I.~O.}\ \bibnamefont
  {Troyanchuk}}, \bibinfo {author} {\bibfnamefont {D.~V.}\ \bibnamefont
  {Karpinsky}}, \bibinfo {author} {\bibfnamefont {M.~V.}\ \bibnamefont
  {Bushinsky}}, \bibinfo {author} {\bibfnamefont {O.~S.}\ \bibnamefont
  {Mantytskaya}}, \bibinfo {author} {\bibfnamefont {N.~V.}\ \bibnamefont
  {Tereshko}}, \ and\ \bibinfo {author} {\bibfnamefont {V.~N.}\ \bibnamefont
  {Shut}},\ }\href {\doibase 10.1111/j.1551-2916.2011.04780.x} {\bibfield
  {journal} {\bibinfo  {journal} {Journal of the American Ceramic Society}\
  }\textbf {\bibinfo {volume} {94}},\ \bibinfo {pages} {4502} (\bibinfo {year}
  {2011})}\BibitemShut {NoStop}%
\bibitem [{\citenamefont {Jeon}\ \emph {et~al.}(2011)\citenamefont {Jeon},
  \citenamefont {Rout}, \citenamefont {Kim},\ and\ \citenamefont
  {Kang}}]{Jeon11}%
  \BibitemOpen
  \bibfield  {author} {\bibinfo {author} {\bibfnamefont {N.}~\bibnamefont
  {Jeon}}, \bibinfo {author} {\bibfnamefont {D.}~\bibnamefont {Rout}}, \bibinfo
  {author} {\bibfnamefont {I.~W.}\ \bibnamefont {Kim}}, \ and\ \bibinfo
  {author} {\bibfnamefont {S.-J.~L.}\ \bibnamefont {Kang}},\ }\href {\doibase
  http://dx.doi.org/10.1063/1.3552682} {\bibfield  {journal} {\bibinfo
  {journal} {Applied Physics Letters}\ }\textbf {\bibinfo {volume} {98}},\
  \bibinfo {eid} {072901} (\bibinfo {year} {2011})}\BibitemShut {NoStop}%
\bibitem [{\citenamefont {Uniyal}\ and\ \citenamefont
  {Yadav}(2009)}]{Uniyal09}%
  \BibitemOpen
  \bibfield  {author} {\bibinfo {author} {\bibfnamefont {P.}~\bibnamefont
  {Uniyal}}\ and\ \bibinfo {author} {\bibfnamefont {K.~L.}\ \bibnamefont
  {Yadav}},\ }\href {http://stacks.iop.org/0953-8984/21/i=1/a=012205}
  {\bibfield  {journal} {\bibinfo  {journal} {Journal of Physics: Condensed
  Matter}\ }\textbf {\bibinfo {volume} {21}},\ \bibinfo {pages} {012205}
  (\bibinfo {year} {2009})}\BibitemShut {NoStop}%
\bibitem [{\citenamefont {Levin}\ \emph {et~al.}(2011)\citenamefont {Levin},
  \citenamefont {Tucker}, \citenamefont {Wu}, \citenamefont {Provenzano},
  \citenamefont {Dennis}, \citenamefont {Karimi}, \citenamefont {Comyn},
  \citenamefont {Stevenson}, \citenamefont {Smith},\ and\ \citenamefont
  {Reaney}}]{Levin11}%
  \BibitemOpen
  \bibfield  {author} {\bibinfo {author} {\bibfnamefont {I.}~\bibnamefont
  {Levin}}, \bibinfo {author} {\bibfnamefont {M.~G.}\ \bibnamefont {Tucker}},
  \bibinfo {author} {\bibfnamefont {H.}~\bibnamefont {Wu}}, \bibinfo {author}
  {\bibfnamefont {V.}~\bibnamefont {Provenzano}}, \bibinfo {author}
  {\bibfnamefont {C.~L.}\ \bibnamefont {Dennis}}, \bibinfo {author}
  {\bibfnamefont {S.}~\bibnamefont {Karimi}}, \bibinfo {author} {\bibfnamefont
  {T.}~\bibnamefont {Comyn}}, \bibinfo {author} {\bibfnamefont
  {T.}~\bibnamefont {Stevenson}}, \bibinfo {author} {\bibfnamefont {R.~I.}\
  \bibnamefont {Smith}}, \ and\ \bibinfo {author} {\bibfnamefont {I.~M.}\
  \bibnamefont {Reaney}},\ }\href {\doibase 10.1021/cm1036925} {\bibfield
  {journal} {\bibinfo  {journal} {Chemistry of Materials}\ }\textbf {\bibinfo
  {volume} {23}},\ \bibinfo {pages} {2166} (\bibinfo {year}
  {2011})}\BibitemShut {NoStop}%
\bibitem [{\citenamefont {Khomchenko}\ \emph {et~al.}(2008)\citenamefont
  {Khomchenko}, \citenamefont {Kiselev}, \citenamefont {Bdikin}, \citenamefont
  {Shvartsman}, \citenamefont {Borisov}, \citenamefont {Kleemann},
  \citenamefont {Vieira},\ and\ \citenamefont {Kholkin}}]{Khomchenko08}%
  \BibitemOpen
  \bibfield  {author} {\bibinfo {author} {\bibfnamefont {V.~A.}\ \bibnamefont
  {Khomchenko}}, \bibinfo {author} {\bibfnamefont {D.~A.}\ \bibnamefont
  {Kiselev}}, \bibinfo {author} {\bibfnamefont {I.~K.}\ \bibnamefont {Bdikin}},
  \bibinfo {author} {\bibfnamefont {V.~V.}\ \bibnamefont {Shvartsman}},
  \bibinfo {author} {\bibfnamefont {P.}~\bibnamefont {Borisov}}, \bibinfo
  {author} {\bibfnamefont {W.}~\bibnamefont {Kleemann}}, \bibinfo {author}
  {\bibfnamefont {J.~M.}\ \bibnamefont {Vieira}}, \ and\ \bibinfo {author}
  {\bibfnamefont {A.~L.}\ \bibnamefont {Kholkin}},\ }\href {\doibase
  http://dx.doi.org/10.1063/1.3058708} {\bibfield  {journal} {\bibinfo
  {journal} {Applied Physics Letters}\ }\textbf {\bibinfo {volume} {93}},\
  \bibinfo {eid} {262905} (\bibinfo {year} {2008})}\BibitemShut {NoStop}%
\bibitem [{\citenamefont {Khomchenko}\ \emph {et~al.}(2010)\citenamefont
  {Khomchenko}, \citenamefont {Paixão}, \citenamefont {Shvartsman},
  \citenamefont {Borisov}, \citenamefont {Kleemann}, \citenamefont
  {Karpinsky},\ and\ \citenamefont {Kholkin}}]{Khomchenko10}%
  \BibitemOpen
  \bibfield  {author} {\bibinfo {author} {\bibfnamefont {V.~A.}\ \bibnamefont
  {Khomchenko}}, \bibinfo {author} {\bibfnamefont {J.~A.}\ \bibnamefont
  {Paixão}}, \bibinfo {author} {\bibfnamefont {V.~V.}\ \bibnamefont
  {Shvartsman}}, \bibinfo {author} {\bibfnamefont {P.}~\bibnamefont {Borisov}},
  \bibinfo {author} {\bibfnamefont {W.}~\bibnamefont {Kleemann}}, \bibinfo
  {author} {\bibfnamefont {D.~V.}\ \bibnamefont {Karpinsky}}, \ and\ \bibinfo
  {author} {\bibfnamefont {A.~L.}\ \bibnamefont {Kholkin}},\ }\href {\doibase
  http://dx.doi.org/10.1016/j.scriptamat.2009.11.005} {\bibfield  {journal}
  {\bibinfo  {journal} {Scripta Materialia}\ }\textbf {\bibinfo {volume}
  {62}},\ \bibinfo {pages} {238 } (\bibinfo {year} {2010})}\BibitemShut
  {NoStop}%
\bibitem [{\citenamefont {Khomchenko}\ \emph
  {et~al.}(2011{\natexlab{a}})\citenamefont {Khomchenko}, \citenamefont
  {Paixão}, \citenamefont {Costa}, \citenamefont {Karpinsky}, \citenamefont
  {Kholkin}, \citenamefont {Troyanchuk}, \citenamefont {Shvartsman},
  \citenamefont {Borisov},\ and\ \citenamefont
  {Kleemann}}]{Khomchenko_0.15_11}%
  \BibitemOpen
  \bibfield  {author} {\bibinfo {author} {\bibfnamefont {V.~A.}\ \bibnamefont
  {Khomchenko}}, \bibinfo {author} {\bibfnamefont {J.~A.}\ \bibnamefont
  {Paixão}}, \bibinfo {author} {\bibfnamefont {B.~F.~O.}\ \bibnamefont
  {Costa}}, \bibinfo {author} {\bibfnamefont {D.~V.}\ \bibnamefont
  {Karpinsky}}, \bibinfo {author} {\bibfnamefont {A.~L.}\ \bibnamefont
  {Kholkin}}, \bibinfo {author} {\bibfnamefont {I.~O.}\ \bibnamefont
  {Troyanchuk}}, \bibinfo {author} {\bibfnamefont {V.~V.}\ \bibnamefont
  {Shvartsman}}, \bibinfo {author} {\bibfnamefont {P.}~\bibnamefont {Borisov}},
  \ and\ \bibinfo {author} {\bibfnamefont {W.}~\bibnamefont {Kleemann}},\
  }\href {\doibase 10.1002/crat.201100040} {\bibfield  {journal} {\bibinfo
  {journal} {Crystal Research and Technology}\ }\textbf {\bibinfo {volume}
  {46}},\ \bibinfo {pages} {238} (\bibinfo {year}
  {2011}{\natexlab{a}})}\BibitemShut {NoStop}%
\bibitem [{\citenamefont {Bielecki}\ \emph {et~al.}(2012)\citenamefont
  {Bielecki}, \citenamefont {Svedlindh}, \citenamefont {Tibebu}, \citenamefont
  {Cai}, \citenamefont {Eriksson}, \citenamefont {B\"orjesson},\ and\
  \citenamefont {Knee}}]{Bielecki12}%
  \BibitemOpen
  \bibfield  {author} {\bibinfo {author} {\bibfnamefont {J.}~\bibnamefont
  {Bielecki}}, \bibinfo {author} {\bibfnamefont {P.}~\bibnamefont {Svedlindh}},
  \bibinfo {author} {\bibfnamefont {D.~T.}\ \bibnamefont {Tibebu}}, \bibinfo
  {author} {\bibfnamefont {S.}~\bibnamefont {Cai}}, \bibinfo {author}
  {\bibfnamefont {S.-G.}\ \bibnamefont {Eriksson}}, \bibinfo {author}
  {\bibfnamefont {L.}~\bibnamefont {B\"orjesson}}, \ and\ \bibinfo {author}
  {\bibfnamefont {C.~S.}\ \bibnamefont {Knee}},\ }\href {\doibase
  10.1103/PhysRevB.86.184422} {\bibfield  {journal} {\bibinfo  {journal} {Phys.
  Rev. B}\ }\textbf {\bibinfo {volume} {86}},\ \bibinfo {pages} {184422}
  (\bibinfo {year} {2012})}\BibitemShut {NoStop}%
\bibitem [{\citenamefont {Nalwa}\ and\ \citenamefont {Garg}(2008)}]{Nalwa08}%
  \BibitemOpen
  \bibfield  {author} {\bibinfo {author} {\bibfnamefont {K.~S.}\ \bibnamefont
  {Nalwa}}\ and\ \bibinfo {author} {\bibfnamefont {A.}~\bibnamefont {Garg}},\
  }\href {\doibase http://dx.doi.org/10.1063/1.2838483} {\bibfield  {journal}
  {\bibinfo  {journal} {Journal of Applied Physics}\ }\textbf {\bibinfo
  {volume} {103}},\ \bibinfo {eid} {044101} (\bibinfo {year} {2008}),\
  http://dx.doi.org/10.1063/1.2838483}\BibitemShut {NoStop}%
\bibitem [{\citenamefont {Maurya}\ \emph {et~al.}(2009)\citenamefont {Maurya},
  \citenamefont {Thota}, \citenamefont {Garg}, \citenamefont {Pandey},
  \citenamefont {Chand},\ and\ \citenamefont {Verma}}]{Maurya09}%
  \BibitemOpen
  \bibfield  {author} {\bibinfo {author} {\bibfnamefont {D.}~\bibnamefont
  {Maurya}}, \bibinfo {author} {\bibfnamefont {H.}~\bibnamefont {Thota}},
  \bibinfo {author} {\bibfnamefont {A.}~\bibnamefont {Garg}}, \bibinfo {author}
  {\bibfnamefont {B.}~\bibnamefont {Pandey}}, \bibinfo {author} {\bibfnamefont
  {P.}~\bibnamefont {Chand}}, \ and\ \bibinfo {author} {\bibfnamefont {H.~C.}\
  \bibnamefont {Verma}},\ }\href
  {http://stacks.iop.org/0953-8984/21/i=2/a=026007} {\bibfield  {journal}
  {\bibinfo  {journal} {Journal of Physics: Condensed Matter}\ }\textbf
  {\bibinfo {volume} {21}},\ \bibinfo {pages} {026007} (\bibinfo {year}
  {2009})}\BibitemShut {NoStop}%
\bibitem [{\citenamefont {Khomchenko}\ \emph
  {et~al.}(2011{\natexlab{b}})\citenamefont {Khomchenko}, \citenamefont
  {Pereira},\ and\ \citenamefont {Paixão}}]{Khomchenko_0.14_11}%
  \BibitemOpen
  \bibfield  {author} {\bibinfo {author} {\bibfnamefont {V.~A.}\ \bibnamefont
  {Khomchenko}}, \bibinfo {author} {\bibfnamefont {L.~C.~J.}\ \bibnamefont
  {Pereira}}, \ and\ \bibinfo {author} {\bibfnamefont {J.~A.}\ \bibnamefont
  {Paixão}},\ }\href {http://stacks.iop.org/0022-3727/44/i=18/a=185406}
  {\bibfield  {journal} {\bibinfo  {journal} {Journal of Physics D: Applied
  Physics}\ }\textbf {\bibinfo {volume} {44}},\ \bibinfo {pages} {185406}
  (\bibinfo {year} {2011}{\natexlab{b}})}\BibitemShut {NoStop}%
\bibitem [{\citenamefont {Dai}\ \emph {et~al.}(2012)\citenamefont {Dai},
  \citenamefont {Chen}, \citenamefont {Li},\ and\ \citenamefont
  {Li}}]{Haiyang12}%
  \BibitemOpen
  \bibfield  {author} {\bibinfo {author} {\bibfnamefont {H.}~\bibnamefont
  {Dai}}, \bibinfo {author} {\bibfnamefont {Z.}~\bibnamefont {Chen}}, \bibinfo
  {author} {\bibfnamefont {T.}~\bibnamefont {Li}}, \ and\ \bibinfo {author}
  {\bibfnamefont {Y.}~\bibnamefont {Li}},\ }\href {\doibase
  http://dx.doi.org/10.1016/S1002-0721(12)60191-4} {\bibfield  {journal}
  {\bibinfo  {journal} {Journal of Rare Earths}\ }\textbf {\bibinfo {volume}
  {30}},\ \bibinfo {pages} {1123 } (\bibinfo {year} {2012})}\BibitemShut
  {NoStop}%
\bibitem [{\citenamefont {Saxin}\ and\ \citenamefont {Knee}(2011)}]{Saxin11}%
  \BibitemOpen
  \bibfield  {author} {\bibinfo {author} {\bibfnamefont {S.}~\bibnamefont
  {Saxin}}\ and\ \bibinfo {author} {\bibfnamefont {C.~S.}\ \bibnamefont
  {Knee}},\ }\href {\doibase 10.1039/C0DT01754J} {\bibfield  {journal}
  {\bibinfo  {journal} {Dalton Trans.}\ }\textbf {\bibinfo {volume} {40}},\
  \bibinfo {pages} {3462} (\bibinfo {year} {2011})}\BibitemShut {NoStop}%
\bibitem [{\citenamefont {Chapon}\ \emph {et~al.}(2011)\citenamefont {Chapon},
  \citenamefont {Manuel}, \citenamefont {Radaelli}, \citenamefont {Benson},
  \citenamefont {Perrott}, \citenamefont {Ansell}, \citenamefont {Rhodes},
  \citenamefont {Raspino}, \citenamefont {Duxbury}, \citenamefont {Spill},\
  and\ \citenamefont {Norris}}]{chapon11}%
  \BibitemOpen
  \bibfield  {author} {\bibinfo {author} {\bibfnamefont {L.~C.}\ \bibnamefont
  {Chapon}}, \bibinfo {author} {\bibfnamefont {P.}~\bibnamefont {Manuel}},
  \bibinfo {author} {\bibfnamefont {P.~G.}\ \bibnamefont {Radaelli}}, \bibinfo
  {author} {\bibfnamefont {C.}~\bibnamefont {Benson}}, \bibinfo {author}
  {\bibfnamefont {L.}~\bibnamefont {Perrott}}, \bibinfo {author} {\bibfnamefont
  {S.}~\bibnamefont {Ansell}}, \bibinfo {author} {\bibfnamefont {N.~J.}\
  \bibnamefont {Rhodes}}, \bibinfo {author} {\bibfnamefont {D.}~\bibnamefont
  {Raspino}}, \bibinfo {author} {\bibfnamefont {D.}~\bibnamefont {Duxbury}},
  \bibinfo {author} {\bibfnamefont {E.}~\bibnamefont {Spill}}, \ and\ \bibinfo
  {author} {\bibfnamefont {J.}~\bibnamefont {Norris}},\ }\href@noop {}
  {\bibfield  {journal} {\bibinfo  {journal} {Neutron News}\ }\textbf {\bibinfo
  {volume} {22}},\ \bibinfo {pages} {22} (\bibinfo {year} {2011})}\BibitemShut
  {NoStop}%
\bibitem [{\citenamefont {Kubota}\ \emph {et~al.}(2011)\citenamefont {Kubota},
  \citenamefont {Oka}, \citenamefont {Nakamura}, \citenamefont {Yabuta},
  \citenamefont {Miura}, \citenamefont {Shimakawa},\ and\ \citenamefont
  {Azuma}}]{Kubota11}%
  \BibitemOpen
  \bibfield  {author} {\bibinfo {author} {\bibfnamefont {M.}~\bibnamefont
  {Kubota}}, \bibinfo {author} {\bibfnamefont {K.}~\bibnamefont {Oka}},
  \bibinfo {author} {\bibfnamefont {Y.}~\bibnamefont {Nakamura}}, \bibinfo
  {author} {\bibfnamefont {H.}~\bibnamefont {Yabuta}}, \bibinfo {author}
  {\bibfnamefont {K.}~\bibnamefont {Miura}}, \bibinfo {author} {\bibfnamefont
  {Y.}~\bibnamefont {Shimakawa}}, \ and\ \bibinfo {author} {\bibfnamefont
  {M.}~\bibnamefont {Azuma}},\ }\href
  {http://stacks.iop.org/1347-4065/50/i=9S2/a=09NE08} {\bibfield  {journal}
  {\bibinfo  {journal} {Japanese Journal of Applied Physics}\ }\textbf
  {\bibinfo {volume} {50}},\ \bibinfo {pages} {09NE08} (\bibinfo {year}
  {2011})}\BibitemShut {NoStop}%
\bibitem [{\citenamefont {Rodr{\'\i}guez-Carvajal}(1993)}]{rodriguezcarvaja93}%
  \BibitemOpen
  \bibfield  {author} {\bibinfo {author} {\bibfnamefont {J.}~\bibnamefont
  {Rodr{\'\i}guez-Carvajal}},\ }\href@noop {} {\bibfield  {journal} {\bibinfo
  {journal} {Physica B}\ }\textbf {\bibinfo {volume} {192}},\ \bibinfo {pages}
  {55} (\bibinfo {year} {1993})}\BibitemShut {NoStop}%
\bibitem [{\citenamefont {Sosnowska}\ \emph {et~al.}(2002)\citenamefont
  {Sosnowska}, \citenamefont {Schäfer}, \citenamefont {Kockelmann},
  \citenamefont {Andersen},\ and\ \citenamefont {Troyanchuk}}]{Sosnowska02}%
  \BibitemOpen
  \bibfield  {author} {\bibinfo {author} {\bibfnamefont {I.}~\bibnamefont
  {Sosnowska}}, \bibinfo {author} {\bibfnamefont {W.}~\bibnamefont {Schäfer}},
  \bibinfo {author} {\bibfnamefont {W.}~\bibnamefont {Kockelmann}}, \bibinfo
  {author} {\bibfnamefont {K.}~\bibnamefont {Andersen}}, \ and\ \bibinfo
  {author} {\bibfnamefont {I.}~\bibnamefont {Troyanchuk}},\ }\href {\doibase
  10.1007/s003390201604} {\bibfield  {journal} {\bibinfo  {journal} {Applied
  Physics A}\ }\textbf {\bibinfo {volume} {74}},\ \bibinfo {pages} {s1040}
  (\bibinfo {year} {2002})}\BibitemShut {NoStop}%
\bibitem [{\citenamefont {Knee}\ \emph {et~al.}(2014)\citenamefont {Knee},
  \citenamefont {Tucker}, \citenamefont {Manuel}, \citenamefont {Cai},
  \citenamefont {Bielecki}, \citenamefont {B\"orjesson},\ and\ \citenamefont
  {Eriksson}}]{Knee14}%
  \BibitemOpen
  \bibfield  {author} {\bibinfo {author} {\bibfnamefont {C.~S.}\ \bibnamefont
  {Knee}}, \bibinfo {author} {\bibfnamefont {M.~G.}\ \bibnamefont {Tucker}},
  \bibinfo {author} {\bibfnamefont {P.}~\bibnamefont {Manuel}}, \bibinfo
  {author} {\bibfnamefont {S.}~\bibnamefont {Cai}}, \bibinfo {author}
  {\bibfnamefont {J.}~\bibnamefont {Bielecki}}, \bibinfo {author}
  {\bibfnamefont {L.}~\bibnamefont {B\"orjesson}}, \ and\ \bibinfo {author}
  {\bibfnamefont {S.~G.}\ \bibnamefont {Eriksson}},\ }\href {\doibase
  10.1021/cm403558j} {\bibfield  {journal} {\bibinfo  {journal} {Chemistry of
  Materials}\ }\textbf {\bibinfo {volume} {26}},\ \bibinfo {pages} {1180}
  (\bibinfo {year} {2014})}\BibitemShut {NoStop}%
\bibitem [{\citenamefont {Kobayashi}\ \emph {et~al.}(2004)\citenamefont
  {Kobayashi}, \citenamefont {Osawa}, \citenamefont {Kimura}, \citenamefont
  {Noda}, \citenamefont {Kasahara}, \citenamefont {Mitsuda},\ and\
  \citenamefont {Kohn}}]{Kobayashi04}%
  \BibitemOpen
  \bibfield  {author} {\bibinfo {author} {\bibfnamefont {S.}~\bibnamefont
  {Kobayashi}}, \bibinfo {author} {\bibfnamefont {T.}~\bibnamefont {Osawa}},
  \bibinfo {author} {\bibfnamefont {H.}~\bibnamefont {Kimura}}, \bibinfo
  {author} {\bibfnamefont {Y.}~\bibnamefont {Noda}}, \bibinfo {author}
  {\bibfnamefont {N.}~\bibnamefont {Kasahara}}, \bibinfo {author}
  {\bibfnamefont {S.}~\bibnamefont {Mitsuda}}, \ and\ \bibinfo {author}
  {\bibfnamefont {K.}~\bibnamefont {Kohn}},\ }\href {\doibase
  10.1143/JPSJ.73.3439} {\bibfield  {journal} {\bibinfo  {journal} {Journal of
  the Physical Society of Japan}\ }\textbf {\bibinfo {volume} {73}},\ \bibinfo
  {pages} {3439} (\bibinfo {year} {2004})}\BibitemShut {NoStop}%
\bibitem [{poi()}]{point_charge}%
  \BibitemOpen
  \href@noop {} {}\bibinfo {note} {A simple point charge approximation
  including the nearest neighbor oxygen sites gives $|A^2_0|$ $>>$ $|A^4_0|$
  and $|A^{4}_{\pm 3}|$. In fact, the dominant $A^2_0$ parameter results in
  non-Ising doublets occupying an almost vanishingly small region of the
  parameter space.}\BibitemShut {Stop}%
\bibitem [{\citenamefont {Sosnowska}\ \emph {et~al.}(2013)\citenamefont
  {Sosnowska}, \citenamefont {Azuma}, \citenamefont {Przeniosło},
  \citenamefont {Wardecki}, \citenamefont {tin Chen}, \citenamefont {Oka},\
  and\ \citenamefont {Shimakawa}}]{sosnowska13}%
  \BibitemOpen
  \bibfield  {author} {\bibinfo {author} {\bibfnamefont {I.}~\bibnamefont
  {Sosnowska}}, \bibinfo {author} {\bibfnamefont {M.}~\bibnamefont {Azuma}},
  \bibinfo {author} {\bibfnamefont {R.}~\bibnamefont {Przeniosło}}, \bibinfo
  {author} {\bibfnamefont {D.}~\bibnamefont {Wardecki}}, \bibinfo {author}
  {\bibfnamefont {W.}~\bibnamefont {tin Chen}}, \bibinfo {author}
  {\bibfnamefont {K.}~\bibnamefont {Oka}}, \ and\ \bibinfo {author}
  {\bibfnamefont {Y.}~\bibnamefont {Shimakawa}},\ }\href {\doibase
  10.1021/ic402427q} {\bibfield  {journal} {\bibinfo  {journal} {Inorganic
  Chemistry}\ }\textbf {\bibinfo {volume} {52}},\ \bibinfo {pages} {13269}
  (\bibinfo {year} {2013})}\BibitemShut {NoStop}%
\bibitem [{\citenamefont {Solovyev}(2012)}]{Solovyev2012}%
  \BibitemOpen
  \bibfield  {author} {\bibinfo {author} {\bibfnamefont {I.~V.}\ \bibnamefont
  {Solovyev}},\ }\href {\doibase 10.1103/PhysRevB.85.054420} {\bibfield
  {journal} {\bibinfo  {journal} {Phys. Rev. B}\ }\textbf {\bibinfo {volume}
  {85}},\ \bibinfo {pages} {054420} (\bibinfo {year} {2012})}\BibitemShut
  {NoStop}%
\bibitem [{\citenamefont {Wang}\ \emph {et~al.}(2011)\citenamefont {Wang},
  \citenamefont {Saal}, \citenamefont {Wu}, \citenamefont {Wang}, \citenamefont
  {Shang}, \citenamefont {Liu},\ and\ \citenamefont {Chen}}]{Wang11}%
  \BibitemOpen
  \bibfield  {author} {\bibinfo {author} {\bibfnamefont {Y.}~\bibnamefont
  {Wang}}, \bibinfo {author} {\bibfnamefont {J.~E.}\ \bibnamefont {Saal}},
  \bibinfo {author} {\bibfnamefont {P.}~\bibnamefont {Wu}}, \bibinfo {author}
  {\bibfnamefont {J.}~\bibnamefont {Wang}}, \bibinfo {author} {\bibfnamefont
  {S.}~\bibnamefont {Shang}}, \bibinfo {author} {\bibfnamefont {Z.-K.}\
  \bibnamefont {Liu}}, \ and\ \bibinfo {author} {\bibfnamefont {L.-Q.}\
  \bibnamefont {Chen}},\ }\href@noop {} {\bibfield  {journal} {\bibinfo
  {journal} {Acta Materialia}\ }\textbf {\bibinfo {volume} {59}},\ \bibinfo
  {pages} {4229 } (\bibinfo {year} {2011})}\BibitemShut {NoStop}%
\bibitem [{\citenamefont {Rovillain}\ \emph {et~al.}(2009)\citenamefont
  {Rovillain}, \citenamefont {Cazayous}, \citenamefont {Gallais}, \citenamefont
  {Sacuto}, \citenamefont {Lobo}, \citenamefont {Lebeugle},\ and\ \citenamefont
  {Colson}}]{Rovillain09}%
  \BibitemOpen
  \bibfield  {author} {\bibinfo {author} {\bibfnamefont {P.}~\bibnamefont
  {Rovillain}}, \bibinfo {author} {\bibfnamefont {M.}~\bibnamefont {Cazayous}},
  \bibinfo {author} {\bibfnamefont {Y.}~\bibnamefont {Gallais}}, \bibinfo
  {author} {\bibfnamefont {A.}~\bibnamefont {Sacuto}}, \bibinfo {author}
  {\bibfnamefont {R.~P. S.~M.}\ \bibnamefont {Lobo}}, \bibinfo {author}
  {\bibfnamefont {D.}~\bibnamefont {Lebeugle}}, \ and\ \bibinfo {author}
  {\bibfnamefont {D.}~\bibnamefont {Colson}},\ }\href {\doibase
  10.1103/PhysRevB.79.180411} {\bibfield  {journal} {\bibinfo  {journal} {Phys.
  Rev. B}\ }\textbf {\bibinfo {volume} {79}},\ \bibinfo {pages} {180411}
  (\bibinfo {year} {2009})}\BibitemShut {NoStop}%
\bibitem [{\citenamefont {Brugger}\ \emph {et~al.}(1993)\citenamefont
  {Brugger}, \citenamefont {Schreiner}, \citenamefont {Roth}, \citenamefont
  {Adelmann},\ and\ \citenamefont {Czjzek}}]{Brugger93}%
  \BibitemOpen
  \bibfield  {author} {\bibinfo {author} {\bibfnamefont {T.}~\bibnamefont
  {Brugger}}, \bibinfo {author} {\bibfnamefont {T.}~\bibnamefont {Schreiner}},
  \bibinfo {author} {\bibfnamefont {G.}~\bibnamefont {Roth}}, \bibinfo {author}
  {\bibfnamefont {P.}~\bibnamefont {Adelmann}}, \ and\ \bibinfo {author}
  {\bibfnamefont {G.}~\bibnamefont {Czjzek}},\ }\href {\doibase
  10.1103/PhysRevLett.71.2481} {\bibfield  {journal} {\bibinfo  {journal}
  {Phys. Rev. Lett.}\ }\textbf {\bibinfo {volume} {71}},\ \bibinfo {pages}
  {2481} (\bibinfo {year} {1993})}\BibitemShut {NoStop}%
\bibitem [{\citenamefont {Henggeler}\ \emph {et~al.}(1998)\citenamefont
  {Henggeler}, \citenamefont {Roessli}, \citenamefont {Furrer}, \citenamefont
  {Vorderwisch},\ and\ \citenamefont {Chatterji}}]{Henggeler98}%
  \BibitemOpen
  \bibfield  {author} {\bibinfo {author} {\bibfnamefont {W.}~\bibnamefont
  {Henggeler}}, \bibinfo {author} {\bibfnamefont {B.}~\bibnamefont {Roessli}},
  \bibinfo {author} {\bibfnamefont {A.}~\bibnamefont {Furrer}}, \bibinfo
  {author} {\bibfnamefont {P.}~\bibnamefont {Vorderwisch}}, \ and\ \bibinfo
  {author} {\bibfnamefont {T.}~\bibnamefont {Chatterji}},\ }\href {\doibase
  10.1103/PhysRevLett.80.1300} {\bibfield  {journal} {\bibinfo  {journal}
  {Phys. Rev. Lett.}\ }\textbf {\bibinfo {volume} {80}},\ \bibinfo {pages}
  {1300} (\bibinfo {year} {1998})}\BibitemShut {NoStop}%
\bibitem [{ora()}]{ora}%
  \BibitemOpen
  \href@noop {} {}\bibinfo {note} {DOI: 10.5287/bodleian:1RRxnoXAP}\BibitemShut
  {NoStop}%
\end{thebibliography}%

\end{document}